\documentclass[journal=jacsat,manuscript=article]{achemso}

\usepackage[version=4]{mhchem}
\usepackage{subcaption}
\usepackage{xcolor}
\usepackage{hyperref}
\usepackage{chemmacros}

\def\affSorbonneU{Sorbonne Université, CNRS, Physico-chimie des Electrolytes et Nanosystèmes Interfaciaux, PHENIX, F-75005, Paris, France}

\author{Emilio Méndez}
\affiliation{\affSorbonneU}
\author{Rocio Semino}
\affiliation{\affSorbonneU}
\email{rocio.semino@sorbonne-universite.fr}

\title{Machine Learning and Molecular Simulations Reveal Mechanisms of ZIFs Polymorph Selection}

\begin{document}

\makeatletter
\setlength\acs@tocentry@height{5.0cm}
\setlength\acs@tocentry@width{5.5cm}
\makeatother

\begin{tocentry}
\includegraphics[width=1\linewidth]{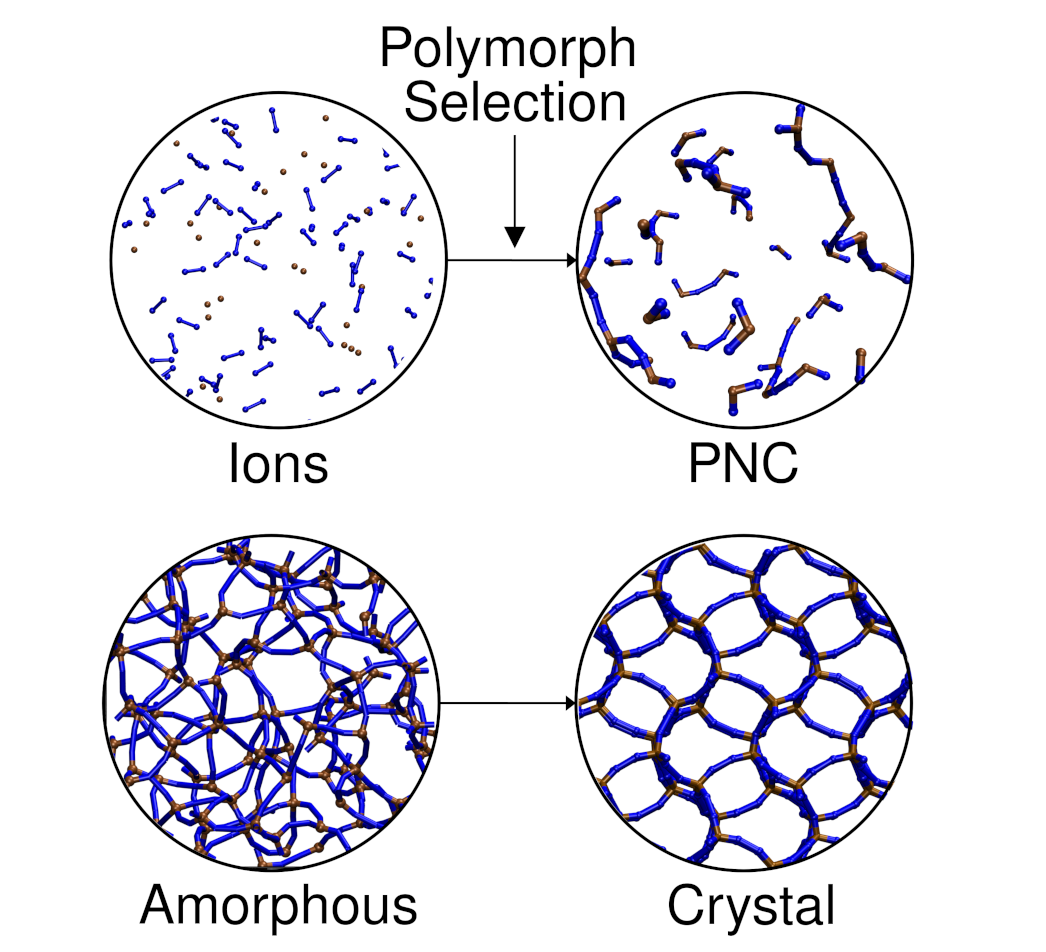}
\end{tocentry}

\begin{abstract}
Zn(imidazolate)$_2$ metal-organic frameworks (MOFs) exhibit a remarkable degree of polymorphism. Because of their promising industrial applications, many research groups have investigated phase transitions, phase diagram and relative stability of these polymorphs. There is now wide consensus in the research community that these MOFs are solvothermally formed via non-classical nucleation mechanisms, in which pre-nucleation clusters are first formed, followed by an intermediate amorphous structure that subsequently reorganizes to yield the final crystalline MOF. However, no study up to date has uncovered which part of the synthesis process determines the final polymorph obtained. In this work, path collective variable metadynamics simulations performed with a partially reactive force field give insights into mechanistic and thermodynamic aspects of the self-assembly of these MOFs. Databases of transient and intermediate synthesis structures are built from the simulations. By developing and applying neural network classifiers over these databases, it is found that both pre-nucleation clusters and the amorphous intermediate structures are polymorph-dependent. These results suggest that polymorph selection happens as early as the pre-nucleation cluster stage.    
\end{abstract}

\section{Introduction}

The 2025 Nobel Prize in Chemistry was awarded for the development of metal-organic frameworks (MOFs), fascinating hybrid materials that possess large cavities that confer them sieving, adsorption and nano-reactor properties. MOF cavities can be tuned by varying the metal and ligand building blocks or by post-synthesis modifications,\cite{Meek2010,He2023} making them suitable for solving a wide range of societal challenges, including water harvesting in deserts,\cite{Kim2018} effective drug delivery\cite{Osterrieth2020} and carbon dioxide capture.\cite{Mahajan2022} Synthesis experts have accumulated decades of experience designing new MOFs for target applications through reticular chemistry principles and interaction-based approaches to choose MOF building blocks.\cite{Yaghi2003,Chen2022,Guillerm2021,Freund2021,Kalmutzki2018,Jiang2021,Stock2011,Guillerm2019,Kitagawa2022} Machine learning methods applied to databases of successful and failed experiments have allowed us to optimize synthesis parameters to form known MOFs.\cite{Moosavi2019} Large language models and self-driven laboratories have provided immense value by using knowledge accumulated by hundreds of groups over more than three decades to propose synthesis protocols for new MOFs.\cite{Zheng2025,Zhao2025,Zheng2023} But despite all this progress, little is known about how these materials form: on the nature of the intermediate species formed during the synthesis and on the steps that connect their formation and disappearance to yield the synthesis product. 

The ZIF-8\cite{Park2006} (Zn(2-methylimidazolate)$_2$) MOF synthesis mechanism has been widely studied in the past two decades by many research groups through a collection of \textit{in situ} and \textit{ex situ} experimental techniques.\cite{Venna2010,Bustamante2014,Balog2022,Moh2013,Cravillon2012,Jin2023,Talosig2024,Dok2025} From this body of works, a global picture of ZIF-8 synthesis has emerged: after a transient period, an amorphous intermediate phase is formed,\cite{Venna2010,Balog2022,Talosig2024,Dok2025} which then reorganizes to give the final crystalline MOF. The work by Talosig and coworkers suggests that this amorphous phase is a critical structure-directing precursor, whose composition and local coordination determine the final crystalline polymorph.\cite{Talosig2024} Prior to the formation of the amorphous intermediate, pre-nucleation clusters (PNCs) are formed.\cite{Talosig2024,Dok2025} PNCs are short-lived and small-sized species, which are difficult to characterize by direct experimental measurements. In the past few years, computer simulation works have shed further light into ZIF synthesis mechanisms.\cite{Filez2021,Balestra2022,Balestra2023,Mendez2025,AndarziGargari2025} PNCs and the amorphous intermediate were also observed \textit{in silico},\cite{Balestra2022} and the structure of the amorphous intermediate was found to be synthesis conditions-dependent (temperature and concentration).\cite{AndarziGargari2025}

The synthesis mechanism of ZIFs formed by the combination of Zn$^{2+}$ and imidazolate (Im$^-$) ions was comparatively much less investigated,\cite{Mendez2025} despite of their rich polymorph landscape.\cite{Park2006,Widmer2019} More than ten Zn(Im)$_2$ polymorphs exist, which can be obtained by varying synthesis conditions (the main variables are metal-to-ligand ratio and temperature)\cite{Park2006} or by exposing a ZIF-4 MOF to higher temperatures or pressures.\cite{Widmer2019} Computational works performed in our group suggest that the synthesis of this kind of ZIFs proceeds \textit{via} a similar global mechanism as that of ZIF-8,\cite{Mendez2025,AndarziGargari2025} with PNCs that grow and merge to form an amorphous intermediate. In analogy with ZIF-8, it could be postulated that polymorph selection will occur at the level of the amorphous intermediate phase.\cite{Talosig2024} Alternatively, polymorph selection could happen prior to the formation of the amorphous phase, at the elusive PNCs stage, or later on, for example \textit{via} an Ostwald ripening mechanism.\cite{Dok2025}

\begin{figure*}[h!]
\centering
\includegraphics[width=0.6\textwidth]{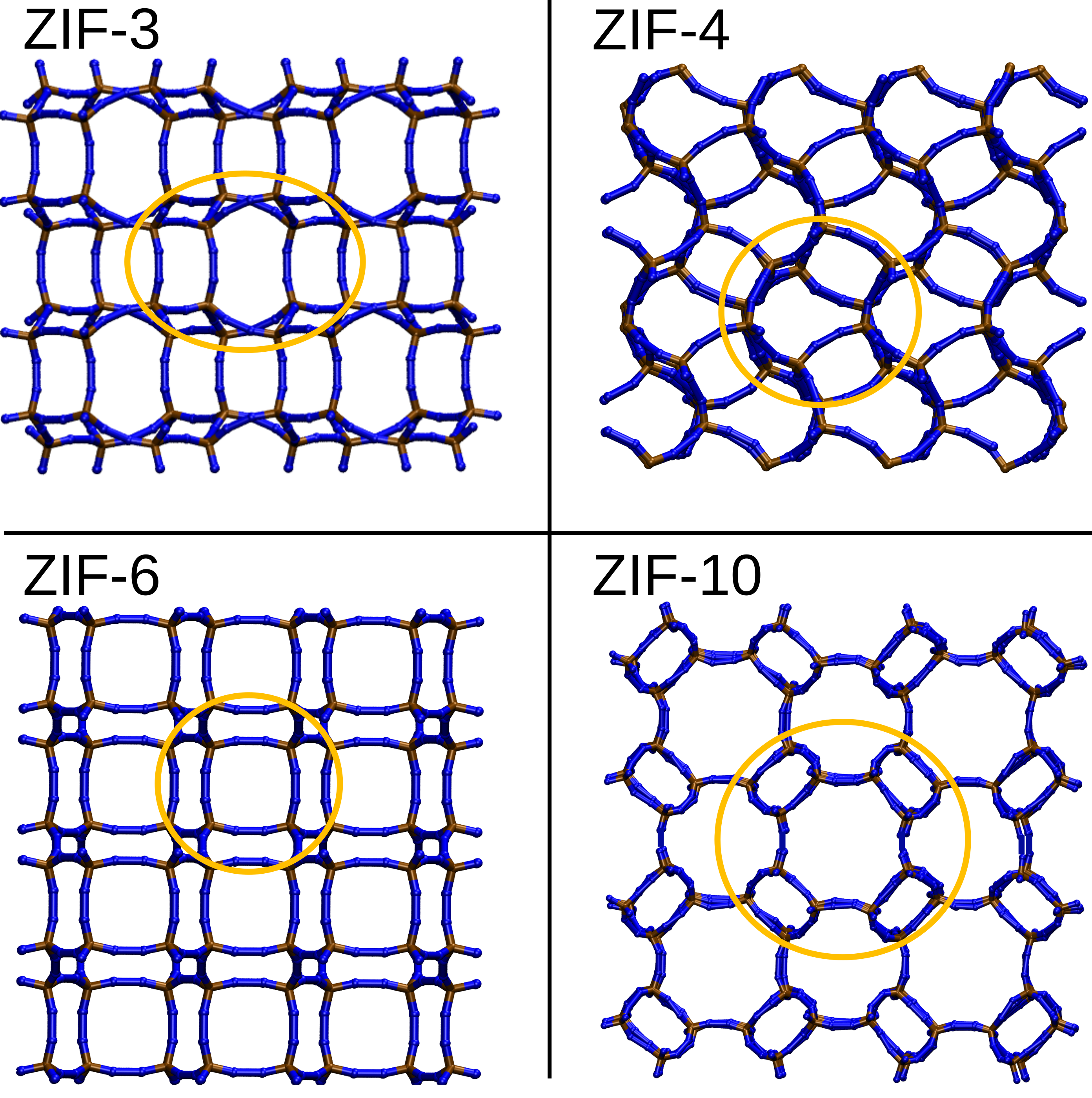}
\caption{\label{fig:snaps_cryst}{Equilibrated crystal structures of the four Zn(Im)$_2$ polymorphs studied in this work, viewed from the XY plane. Ochre vertices and blue edges correspond to Zn and N atoms of the same ligand respectively. For clarity purposes, all other ligand atomic species are not displayed. The yellow circles denote the pores that are used as target structures in Section 2.2.}}
\end{figure*}

In this work, we combine computer simulation and machine learning methods to tackle a central question in elucidating the synthesis mechanism of ZIFs: at which stage of the synthesis process does polymorph selection occur? We model the ZIFs through the nb-ZIF-FF partially reactive force field,\cite{Balestra2022} that has been validated for a wide range of ZIF polymorphs, and we use adaptive path collective variable metadynamics\cite{Laio2002,DazLeines2012} simulations to obtain a statistically representative set of configurations of both the amorphous intermediates and PNCs for a series of Zn(imidazolate)$_2$ polymorphs (ZIF-3, ZIF-4, ZIF-6 and ZIF-10, see \textbf{Figure \ref{fig:snaps_cryst}}). Then, we compare structures associated to different polymorphs employing neural network classifiers,\cite{Mendez2024} to determine whether they are distinguishable or not. Our results suggest that the structure of PNCs is already polymorph-dependent, thus leading us to hypothesize that polymorph selection occurs at synthesis stages as early as the formation of PNCs, even before the amorphous intermediate is formed.  

\section{Results}

We aim to discover in which stage of the self-assembly process occurs polymorph selection for Zn(Im)$_2$ ZIFs. Our underlying hypothesis is that all transient species that are formed along the self-assembly process will be indistinguishable for all polymorphs until the stage in which polymorph selection takes place. From then on, the species are expected to structurally diverge until the final (metastable) phase that will constitute the synthesis product is formed. In principle, polymorph selection could happen either at the PNCs, amorphous intermediate or at later nucleation or growth stages. Therefore, we seek to compare the structures of PNCs and amorphous intermediates that are formed in the synthesis process of a series of polymorphs (see Figure~\ref{fig:snaps_cryst}) by setting up two series of simulation experiments. One is dedicated to obtaining and comparing amorphous intermediate structures while the other set samples PNCs for all polymorphs studied.    

\subsection{Amorphous intermediates}

\begin{figure*}[h!]
\centering
\includegraphics[width=0.9\textwidth]{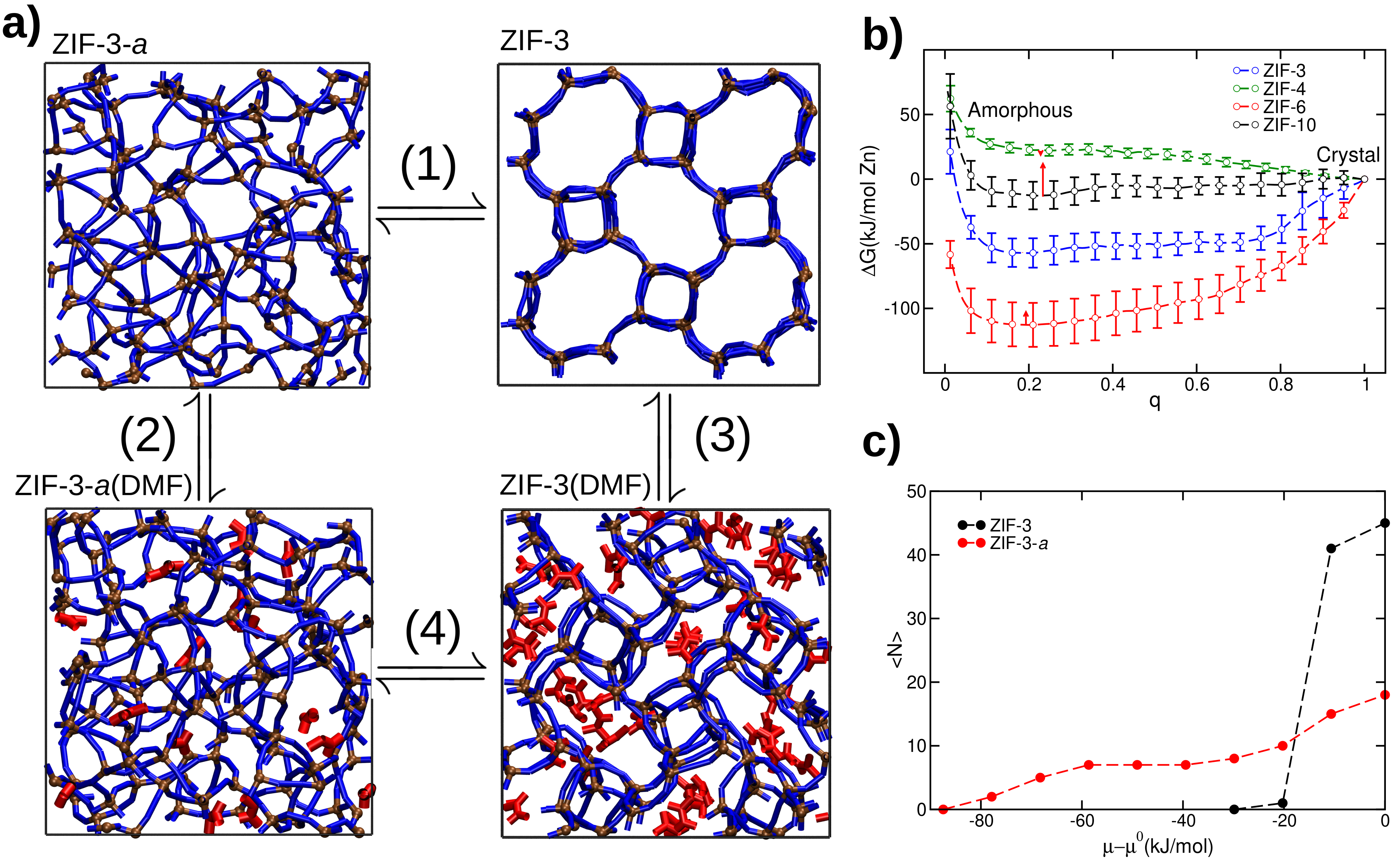}
\caption{\label{fig:amorph_tot}{a) Thermodynamic cycle to study the amorphous-to-crystalline transition in solution, exemplified by ZIF-3. DMF solvent molecules are coloured in red while the rest of the colours correspond to those in Figure \ref{fig:snaps_cryst}. b) Free energy profiles for reaction (1) for the four polymorphs studied. Results are aligned so that the crystal state corresponds to the zero of free energy. c) DMF adsorption curves for ZIF-3 and ZIF-3-\rm{a} as a function of the chemical potential $\mu$. $\mu^0$ corresponds to the chemical potential of liquid DMF at $T=400$ K and $P=1$ bar.}}
\end{figure*}

Since the studies by Talosig \textit{et al.} on ZIF-8 have shown that the structure of the amorphous intermediates leading to different Zn(2-mIm)$_2$ polymorphs are distinct,\cite{Talosig2024} we first focused on these species to see whether this is also the case for Zn-Im-based ZIFs. This requires carrying out independent simulation experiments to obtain amorphous structures for each polymorph. Moreover, a large set of microstructures for each amorphous intermediate is needed to have statistically significant results. To guarantee the correlation between the amorphous intermediate and the crystalline form, we produced the amorphous structures from the void (activated) crystals, by melting and quenching them, and then optimized the thermodynamic path between initial and final states. This also allows us to recover thermodynamic information of the amorphous-to-crystal transition. To do this, an optimal reaction path that goes from amorphous (reactant state) to crystalline (product state) phases was obtained by an adaptive path collective variable scheme detailed in the Methods Section and in the Supporting Information (SI), in which a single collective variable (CV) is obtained from a high-dimensional space of features.\cite{DazLeines2012} An example of the reaction considered is labelled as (1) in panel a) of \textbf{Figure \ref{fig:amorph_tot}}, for the case of ZIF-3.

The degree of advance along the optimal, converged path defined a reaction coordinate $q$, which takes the value of 0 for reactants and 1 for products. The variable $q$ was then used as a CV within a metadynamics\cite{Laio2002} simulation in order to reconstruct the underlying free energy landscape $\Delta$G($q$) and obtain statistically meaningful data, taking into account the stochastic nature of the crystallization process.
The resulting free energies are plotted in panel b) of Figure \ref{fig:amorph_tot}.

The first thing we can note is that the minima that correspond to the amorphous intermediates are not located at $q=$ 0. This means that upon optimizing the path, a thermodynamically more stable amorphous phase was discovered by the algorithm. This is a good sign, as the initial species we proposed for $q=$ 0 was the result of a single melt-quenching trajectory, and thus did not contemplate other possible pathways. The optimization of the path allows us to recover the minimum energy amorphous state, with values that vary for each polymorph but that are always around $q\sim0.18$. Moreover, none of the free energy plots presents significant activation barriers. This indicates that the amorphous-to-crystal transitions are mainly limited by the slow kinetics of Zn--N bond breaking and formation and solvent rearrangements. 

Furthermore, we observe that the relative stability of each amorphous state with respect to the corresponding crystal significantly varies between the different polymorphs. 
In these thermodynamic conditions, the ZIF-4 crystal is more stable than its amorphous counterpart. For ZIF-10, the free energy difference lies inside the error bars, suggesting that the two states are close to equilibrium at these thermodynamic conditions. For ZIF-3 and ZIF-6, the amorphous state is more stable than the crystalline form, which suggests that the crystallization in vacuum is not spontaneous within the thermodynamic conditions studied.
Table 2 of the SI summarizes the synthesis conditions in which each of the polymorphs were experimentally obtained.\cite{Park2006} 
Remarkably, these results match an experimentally observed tendency: only ZIF-4 and ZIF-10 polymorphs were synthesized and activated under these thermodynamic conditions. Conversely, ZIF-3 synthesis requires more specific synthesis conditions, such as a combination of solvents (N,N-Dimethylformamide --DMF-- and N-methylpyrrolidinone) and ZIF-6 was only obtained by combinatory experiments, and it is believed that its structure collapses upon solvent removal.\cite{BousselduBourg2014}
This suggests that the solvent, which in all synthesis experiments was DMF,\cite{Park2006} may have a major influence in the stability of the crystal structures. Indeed, more compact structures are favoured by the increase in ligand-ligand interactions, but structures with larger pore sizes are stabilized by solvent adsorption.
The pore sizes of the four studied polymorphs follow this order from largest to smallest: ZIF-10 (12.2 \AA) $>$ ZIF-6 (8.8 \AA) $>$ ZIF-3 (8.0 \AA) $>$ ZIF-4 (2.0 \AA).

Taking into account the effect of solvent in the reaction is not straightforward, since the amount of solvent molecules adsorbed in the amorphous and crystalline phases is not the same.
To overcome this problem, we considered a thermodynamic cycle of four states, as depicted in panel a) of Figure \ref{fig:amorph_tot}. Reaction (1) corresponds to the amorphous-to-crystal transition in vacuum, for which we computed the free energy \textit{via} the path collective variable scheme discussed above. Reactions (2) and (3) correspond to the filling of the amorphous and crystalline phases with DMF respectively. To compute the free energies associated with these reactions, we performed hybrid Grand Canonical Monte Carlo (GCMC)/molecular dynamics (MD) simulations in the Osmotic Ensemble, in which the systems are exposed to a solvent reservoir at a fixed chemical potential until equilibrium is reached.\cite{Coudert2011,Rogge2019_2,Jeffroy2008} The effect of volume expansion during guest adsorption is taken into account, since the box dimensions are allowed to relax during the MD steps of the algorithm. 
Details of these simulations are given in the SI.
Finally, the free energy of reaction (4), which corresponds to the amorphous-to-crystal transition in presence of the solvent, can be obtained by combining the results of reactions (1), (2) and (3), since free energy is a state function.

The adsorption curves for ZIF-3 and the corresponding amorphous phase ZIF-3-\textit{a} are shown in Figure \ref{fig:amorph_tot} c) as an example. The curves for the rest of the polymorphs present similar features and are displayed in the SI.
They are computed as the average amount of adsorbed solvent molecules $\left\langle{N}\right\rangle$ as a function of the chemical potential of the solvent $\mu$. The reference $\mu^0$ depicts the chemical potential of liquid DMF at $T=$ 400 K and $P=$ 1 bar. The final adsorption free energies are computed from the integral of these plots.
The results, which are summarized in \textbf{Table \ref{adsorption}}, show that the crystalline forms are capable of adsorbing more solvent molecules than the amorphous counterparts due to their higher porosity. However, solvent--framework interactions are stronger in the amorphous phase, since the defective Zn moieties can serve as binding sites. This can be observed from the fact that solvent molecules remain adsorbed to the amorphous phase at lower values of $\mu$ than for the crystal.
As a consequence, the solvent contribution to the reaction free energy can be positive or negative for different polymorphs. For ZIF-4, the solvent stabilizes the amorphous phase, giving a $\Delta \Delta G$ between the reaction with and without solvent of +3.1 kJ/mol Zn. For ZIF-3, both effects almost compensate, leading to a $\Delta \Delta G$ = +0.2 kJ/mol Zn. For ZIF-6 and ZIF-10, which present larger pore sizes, the solvent predominantly contributes to the stabilization of the crystalline form, with values of $\Delta \Delta G$ of -9.8 kJ/mol Zn and -25.5 kJ/mol Zn, respectively. The solvent corrections to the free energies of the reactions are depicted in Figure \ref{fig:amorph_tot} b) with red arrows. In the case of ZIF-10, the solvent makes the amorphous-to-crystal transition spontaneous at the thermodynamic conditions studied.
Overall, our results show that the adsorption free energies of the amorphous phases are similar for all the polymorphs while the adsorption free energies of the crystalline phases are correlated with pore size. The amorphous phases curves have two steps: the first plateau, which corresponds to the adsorption of solvent molecules in the defective sites of the structure, and a second one at values of $\mu-\mu^0\sim -20$ kJ/mol, similar to the ones where adsorption peaks for the crystals. This second increase in adsorption corresponds to the condensation of solvent clusters inside the structure. 

\begin{table}[h]
\centering
\small
  \caption{Free energies for all the reaction steps of the cycle depicted in Figure \ref{fig:amorph_tot} panel a) for each polymorph in kJ/mol Zn. In reactions (1) and (4), crystalline and amorphous phases are considered as product and reactant states, respectively. In reactions (2) and (3) solvent filled forms are considered as product states and empty frameworks as reactants.}
  \label{adsorption}
  \begin{tabular*}{0.5\textwidth}{@{\extracolsep{\fill}}lllll}
     & $\Delta$G$^{(1)}$ & $\Delta$G$^{(2)}$ & $\Delta$G$^{(3)}$ & $\Delta$G$^{(4)}$\\
    \hline
ZIF-3  & +57.7  & -5.4 &   -5.2  &  57.9 \\
ZIF-4  & -22.7  & -3.7 &   -0.8  &  -25.6 \\
ZIF-6  & +112.8 & -4.6 &   -14.0 &  103.4 \\
ZIF-10 & +12.8  & -6.7 &   -32.4 &  -12.9 \\
\hline
  \end{tabular*}
\end{table}

Our thermodynamic analyses allow us to assess relative thermodynamic stability between amorphous and crystal phases, but they do not allow us to determine whether the amorphous phases that correspond to the different polymorphs are equal or not, since the free energies of the crystals are different, so no comparison between the amorphous phases can be made from Figure \ref{fig:amorph_tot}. To asses whether they differ or not, we performed a quantitative comparison of the structural properties of the different amorphous phases instead. As mentioned above, Figure \ref{fig:amorph_tot} b) shows that the minima that correspond to the amorphous phases are not located at $q=$ 0, which represents the initial melt-quenched structure, but instead are found at higher values.
We extracted all the configurations that lie in the vicinity of the local minima from the metadynamics trajectories.
By visual inspection, the configurations look amorphous, see for example ZIF-3-\rm{a} in Figure \ref{fig:amorph_tot} a), and the Zn--N $g(r)$ is also equivalent for the four amorphous species, as shown in Figure S9 of the SI.

\begin{figure*}[h!]
\centering
\includegraphics[width=0.4\textwidth]{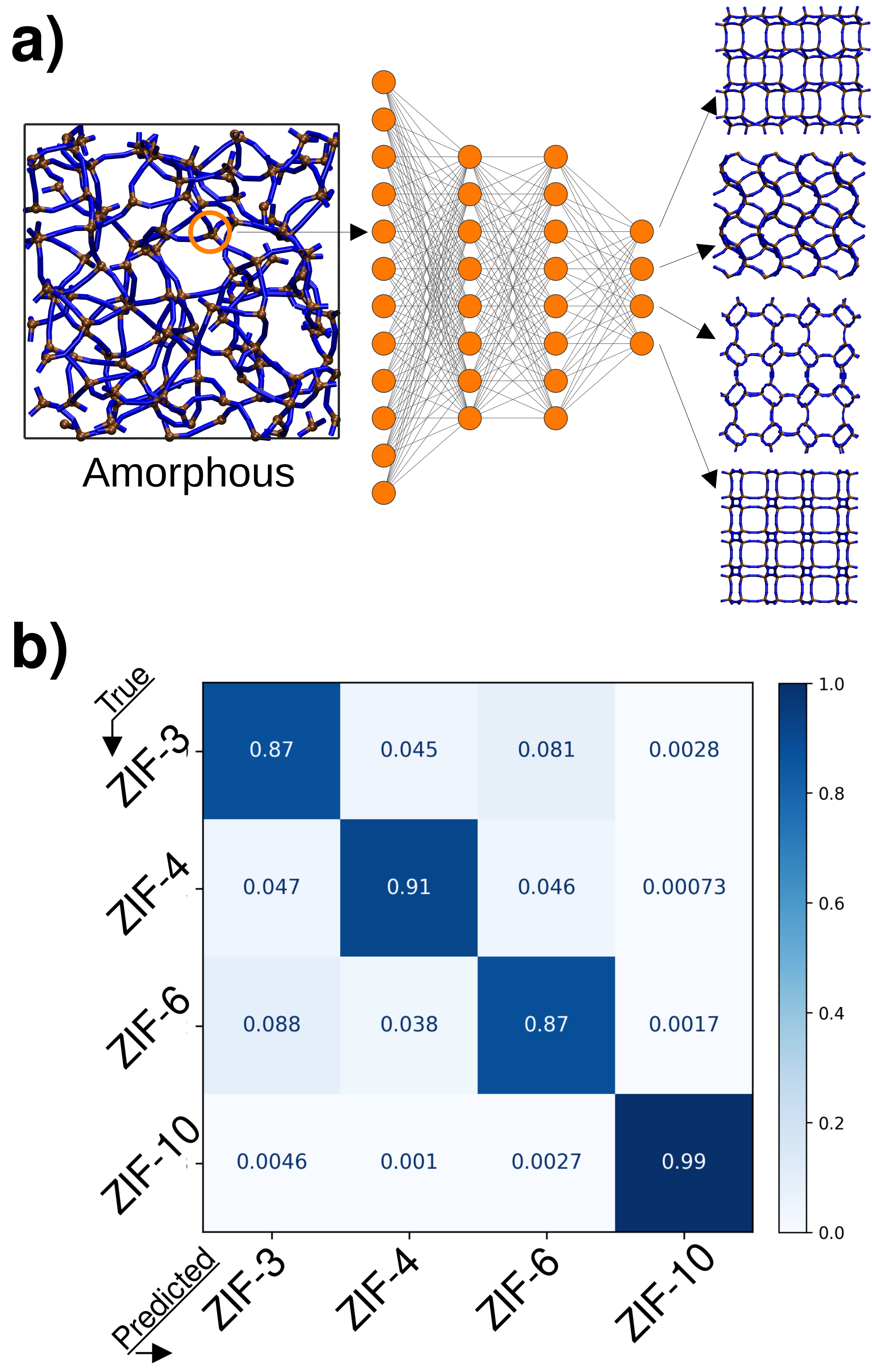}
\caption{\label{fig:confusion_amorph}{a) Scheme of the employed classification algorithm: a Zn-centred environment is described through a set of Behler-Parrinello symmetry functions,\cite{Behler2011} which serve as inputs to a neural network. The outputs of the neural network represent the probabilities of the input environment to come from each of the polymorphs. b) Confusion matrix of the classification algorithm trained to distinguish the four amorphous phases.}}
\end{figure*}

To perform a more detailed comparison, we developed a neural network classification algorithm trained to distinguish between Zn local environments that come from the four amorphous phases, as shown in \textbf{Figure \ref{fig:confusion_amorph}} a), following the same methodology we developed in previous work.\cite{Mendez2024} Details of this procedure are provided in the Methods Section and in the SI. If the algorithm that is trained over structural features (Behler-Parrinello symmetry functions\cite{Behler2011}) is capable of classifying the environments correctly, then the phases are structurally different. Figure \ref{fig:confusion_amorph} b) shows the confusion matrix of the resulting algorithm applied to the test data set. In this matrix, diagonal values show correct predictions, while non-diagonal values quantify the error fraction for the corresponding (real,predicted) pair. The overall accuracy of the classifier is of 93.5\%, which implies that the local environments are effectively different. Moreover, phases with similar crystal pore sizes, particularly ZIF-3 and ZIF-6 (8.0 and 8.8 \AA \ respectively) present the highest classification error, suggesting that the amorphous phases could already exhibit traces of the porosity of the crystalline phases.

To give a physical interpretation to the neural network classifier results and obtain insights into the crystallization mechanism, we measured a set of geometrical parameters as a function of the reaction coordinate $q$. In particular, we monitored the average amount of 3-coordinated Zn$^{2+}$ ions (Zn-3), which is known to play an important role in the crystal-to-amorphous transition of ZIF-4.\cite{Mendez2024} We also measured the average amount of 3-, 4-, 5- and 6-membered rings of Zn$^{2+}$ ions connected by ligands, which is also an indicator of the degree of amorphization, since the crystal structures only contain 4-, 6- and 8-membered rings.\cite{AndarziGargari2025}
The results are shown in Figure S10 in the SI for all the polymorphs. 
All crystallization transitions are accompanied by an increase in the amount of Zn-3 species, followed by a subsequent decrease. This corresponds to the rearrangement of the amorphous defective structure to yield the final crystal, in which all Zn$^{2+}$ ions are 4-coordinated. The amount of Zn-3 at the amorphous state is around 20 for the four polymorphs, which corresponds to $\sim$15\% of the total amount of Zn$^{2+}$ ions. We note that the average amount of Zn-3 does not decay to zero in the crystalline state for all the polymorphs, as would be expected. Instead, a certain amount of defective sites remains present in the crystals: $\sim$ 4\% for ZIF-4 and ZIF-10, $\sim$ 8\% for ZIF-3 and $\sim$ 16\% for ZIF-6. This trend correlates with the free energy values of Figure \ref{fig:amorph_tot} b). It seams that the structures that are less stable, in particular ZIF-6, are not able to form without the presence of defects. From the thermodynamic point of view, although defects are energetically unfavourable, they can be entropically stabilized, and both energetic and entropic effects are taken into account in the simulations. It is worth to mention that these values correspond to average results, and that the non-defective states are also formed during the simulations.
Moreover, the amorphous phases present significant differences in terms of their $n$-membered rings: while ZIF-4-\rm{a} presents mostly 4- and 6-membered rings and a small amount of 3- and 5-membered rings, ZIF-3-\rm{a} and ZIF-6-\rm{a} present comparable amounts of 3-, 4-, 5- and 6-membered rings. Finally, ZIF-10-\rm{a} is rich in 4-, 5- and 6-membered rings but poor in 3-membered rings. This supports the fact that the amorphous states are not equivalent for all the polymorphs. These structural differences are captured by the classification algorithm.

\subsection{Pre-nucleation clusters}

In the previous section, we found that the amorphous intermediate phases formed during the synthesis of a series of ZIF polymorphs present structural differences, which leads us to hypothesize that polymorph differentiation could occur before or at the stage of amorphous intermediate formation.
In this section, we compare pre-nucleation clusters for the polymorphs, to see whether they are structurally different or not. As for the amorphous intermediates, independent simulations must be carried out for each polymorph to extract a statistically meaningful amount of microstates compatible with the PNCs for each of them. To this end, we set up a series of simulations where a pore of each of the polymorphs under consideration was formed starting from a solution of Zn$^{2+}$ and Im$^-$ ions in DMF. 

\begin{figure*}
\centering
\includegraphics[width=0.4\textwidth]{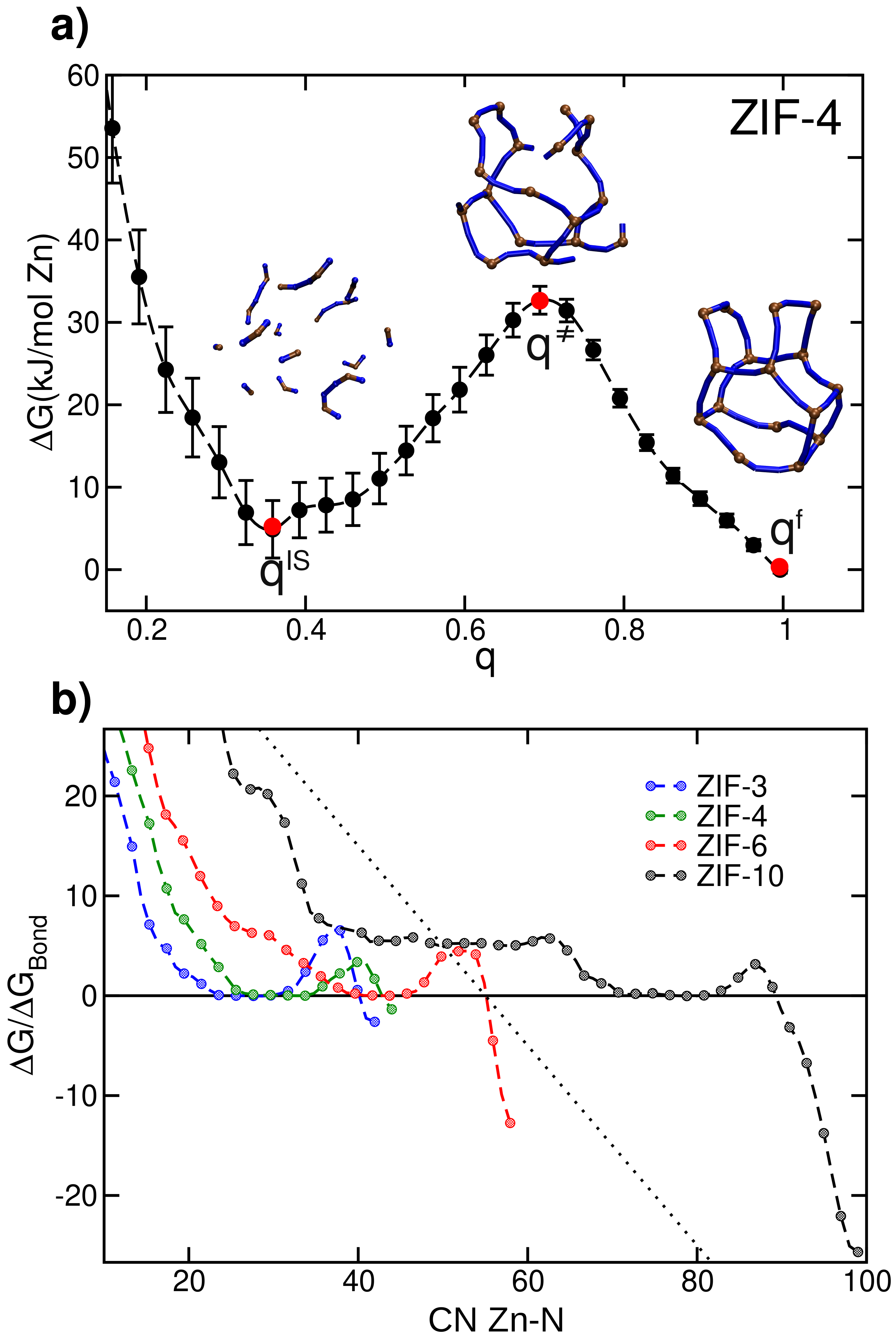}
\caption{\label{fig:free_energy}{a) Free energy per Zn$^{2+}$ ion as a function of the reaction coordinate $q$ for ZIF-4 pore formation. $q$ values for the intermediate state ($q^{\mathrm{IS}}$), the transition state ($q^{\transitionstatesymbol}$) and the final state ($q^\mathrm{f}$=1) are displayed in red. Representative snapshots of the formed cluster in these three stages are included, following the color code in Figure \ref{fig:snaps_cryst}.
b) Free energy as a function of number of Zn--N bonds formed (CN) for each polymorph. The free energies are normalized by the free energy of Zn--N bond formation for an isolated Zn--Im pair in solution.\cite{Mendez2025} In order to make a clear comparison between activation energies, the curves were aligned in such a way that the intermediate state has a free energy of $\Delta$G=0 in all cases. 
The dotted diagonal curve corresponds to the free energy that would be observed if all bonds contributed the same free energy drop $\Delta$G$_{\mathrm{bond}}$.}}
\end{figure*}

For each of the polymorphs considered, a pore formed by connected Zn$^{2+}$ and Im$^-$ ions was cut from the crystal structure, as shown in Figure \ref{fig:snaps_cryst}. 
An optimal reaction path that goes from the solvated ions in a DMF solution (reactant state) to the target pore (product state) was obtained following the same adaptive path collective variable procedure as described above. During the pore formation reaction, the concentration of Zn$^{2+}$ and Im$^-$ ions was kept constant in a spherical slab outside the spherical reaction region using the constant chemical potential molecular dynamics method (C$\mu$MD) to avoid reactant depletion as the pore is formed.\cite{AndarziGargari2026,Karmakar2019}
The degree of advance in the path $q$ was further used as CV within a metadynamics\cite{Laio2002} simulation to reconstruct the underlying free energy landscape $\Delta$G($q$), as for the amorphous intermediates analyses.
The resulting free energy is plotted in panel a) of \textbf{Figure \ref{fig:free_energy}} for ZIF-4, as an example. 

The free energy profile presents two minima, one at $q=$1 that corresponds to the formed pore, and another, higher in energy, that corresponds to an intermediate state, at a value of $q^{\rm{IS}}$=0.37. Both minima are separated by a transition state located at the maximum at $q^{\transitionstatesymbol}$=0.7.
 Three typical snapshots of the system at each of these states were included in the free energy plot of Figure \ref{fig:free_energy}. It is worth remembering that the manifold of configurations that corresponds to a given $q$ value is highly degenerated due to entropic effects. As a consequence, a large amount microstates are compatible with each $q$ value. To show this structural diversity, more snapshots of the PNCs associated to each polymorph are included in the SI.
 
The shape of the free energy plot indicates that pore formation takes place in two steps. The first step, that goes from $q=$0 to $q=$$q^{\rm{IS}}$, involves the transition between fully disconnected, solvated ions to oligomeric clusters that comprise a small amount of Zn$^{2+}$ and ligands ions, as observed in the snapshot for q$^{\rm{IS}}$. These metastable oligomeric clusters can be assimilated to the PNCs, which were experimentally found within the mechanism of self-assembly of ZIF-8.\cite{Dok2025} PNCs are usually found in non-classical nucleation mechanisms in which the binding energy of individual monomers is negative, contrary to the classical nucleation case. This means that the system does not need to reach a critical nucleus size to grow spontaneously. In previous work, we proved that the free energy of Zn--Im binding in solution is negative, which supports the PNC mechanism.\cite{Mendez2025} PNC were also observed by Andarzi Gargari and Semino in a computational study were the synthesis mechanism of ZIF-8 and a Zn-Im-based ZIF were modelled.\cite{AndarziGargari2025}The second step, that goes from $q=$$q^{\rm{IS}}$ to $q=$1, involves further addition of Zn$^{2+}$ and ligand ions to the PNCs and their merging through random connexions to form a bigger structure. Subsequently, this structure undergoes a connectivity rearrangement to achieve the specific topology of the target pore. This rearrangement, along with the concomitant solvent displacements, is responsible for the rise in G($q$) until $q$=$q^{\transitionstatesymbol}$. At a later stage in which the connectivity is almost at the desired final, target value, the final pore is obtained by the formation of a few missing bonds, which are already suggested in the transition state configuration, as can be observed in the snapshot for $q^{\transitionstatesymbol}$. Subsequently, there is a drop in free energy that compensates the rise that takes place in the formation of the transition state. Since the topology rearrangement that leads to the final crystal involves the breaking and formation of several bonds, displacement of internal solvent molecules and a reduction of entropy, modelling it requires timescales much larger than typical simulations times.\cite{Neha2022}
In addition, the small size of the pores considered herein limits the information that can be extracted for the amorphous-to-crystal transition in solution. Previous results from our group have shown that amorphous states can comprise hundreds of Zn$^{2+}$ and ligand ions,\cite{AndarziGargari2025,Balestra2022} while in this case only small clusters comprising from 36 to 60 ions are considered, depending on the polymorph. 

\begin{figure*}[h!]
\centering
\includegraphics[width=0.6\textwidth]{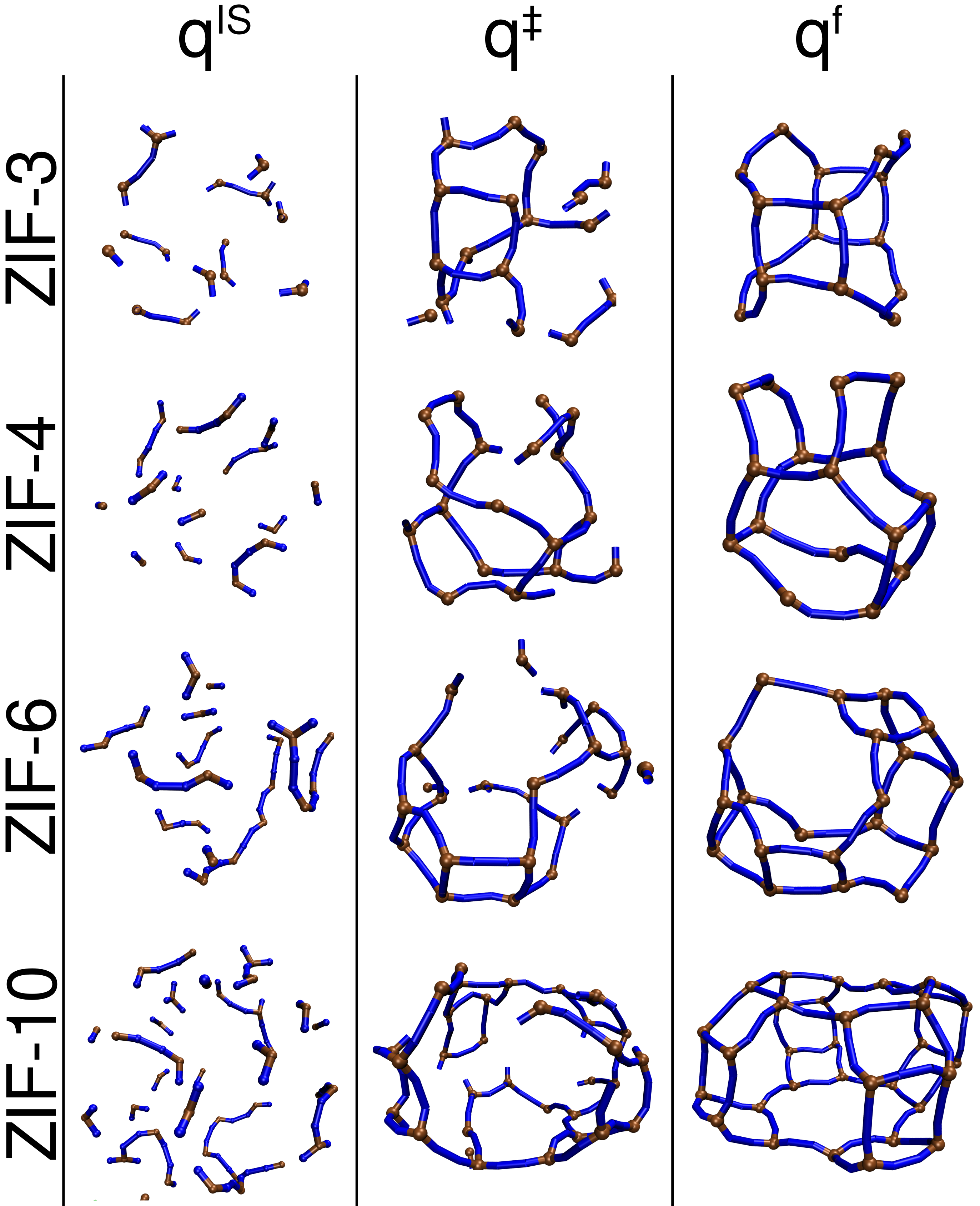}
\caption{\label{fig:snaps}{Representative snapshots of the obtained structures along the pore formation metadynamics simulations for each polymorph for different $q$ values. Only ions with connectivity greater than zero are displayed. $q^{\mathrm{IS}}$ corresponds to the intermediate state, $q^{\transitionstatesymbol}$} to the transition state and $q^\mathrm{f}$ to the global minimum state, which represents a pore defined as shown in Figure \ref{fig:snaps_cryst}.}
\end{figure*}

As mentioned above, the same procedure was carried out for ZIF-3, ZIF-6 and ZIF-10 as well. Their corresponding free energy plots present similar qualitative features and are included in the SI. Typical configurations of each of them at key $q$ values are summarized in \textbf{Figure \ref{fig:snaps}}. 

A direct comparison between the free energy profiles shown in Figure \ref{fig:free_energy} and Figure S11 is not possible since the amount of Zn--N bonds that need to be formed to yield the target pore differs for each polymorph.
For that reason, and to use a variable with physicochemical meaning, we plotted the free energy as a function of the total coordination number between Zn$^{2+}$ and Im$^-$ (CN) instead of $q$. CN represents the total amount of Zn--Im bonds. This change of variables was performed through a standard reweighting procedure explained in the SI. The results are shown in panel b) of Figure~\ref{fig:free_energy}. Within this representation, we can compare the free energy cost of forming a given amount of bonds between polymorphs. The free energies are aligned in such a way that the intermediate state that corresponds to the PNCs is assigned $\Delta$G=0 for a better comparison. Results are normalized by the magnitude of the Zn--N bond formation free energy of an isolated Zn--Im pair in solution, which corresponds to 34.8 kJ/mol for this model.\cite{Mendez2025} A diagonal line of slope -1 is also included in the plot as a reference. If all bonds formed would contribute the same drop in free energy as in the isolated Zn--Im system, then the plots would follow this diagonal line. As expected, all the plots still present two minima, with the exception of the one that corresponds to ZIF-10, which presents a third minimum in the region of CN between 50 and 60. This minimum can also be observed in the plot of $\Delta$G($q$) but it is amplified when the free energy is represented as a function of CN. The plot corresponding to ZIF-10 was aligned to attribute the zero of $\Delta$G to the second minimum, because it presents the highest associated activation energy.

To gain insights in the polymorph selection process that takes place during ZIF synthesis, we should compare pore formation activation energies $\Delta$G$^{\transitionstatesymbol}$, since these are directly related to the relative rate of formation. These correspond to the free energy differences between intermediate and transition states in the plots in Figure \ref{fig:free_energy}, panel b). The results are summarized in \textbf{Table \ref{activation}}. 
The following trend was obtained:  $\Delta$G$_{\rm{ZIF}-4}^{\transitionstatesymbol}$ $\simeq$ $\Delta$G$_{\rm{ZIF}-10}^{\transitionstatesymbol}$ $<$ $\Delta$G$_{\rm{ZIF}-6}^{\transitionstatesymbol}$ $<$ $\Delta$G$_{\rm{ZIF}-3}^{\transitionstatesymbol}$. 
For ZIF-4 and ZIF-10, the order of magnitude of the activation energies is comparable to those reported in experimental studies for similar systems. For example, ZIF-8 has an activation energy of $\sim$100 kJ/mol.\cite{Moh2013,VanVleet2018} Higher activation energies, like the ones computed for ZIF-3 and ZIF-6, suggest that those processes have an extremely slow kinetics within our model.

\begin{table}[h]
\centering
\small
  \caption{Activation free energies of pore formation reactions. }
  \label{activation}
  \begin{tabular*}{0.3\textwidth}{@{\extracolsep{\fill}}ll}
   ZIF  & $\Delta$G$^{\transitionstatesymbol}$ (kJ/mol)  \\
    \hline
ZIF-3  &  205$\pm$ 20\\
ZIF-4  &  104$\pm$ 18\\
ZIF-6  &  149$\pm$ 22\\
ZIF-10 & 116$\pm$ 20\\
\hline
  \end{tabular*}
\end{table}

The activation energy value is the result of a complex interplay between enthalpy, that tends to decrease $\Delta$G$^{\transitionstatesymbol}$ due to the rise in the amount of bonds, and entropy, that tends to increase $\Delta$G$^{\transitionstatesymbol}$ due to a reduction in the degeneracy of microstates as the pore forms.\cite{Mendez2024} The solvent also has an impact, as larger pores allow more solvent molecules to be adsorbed, which results in a higher relative stability. On the other hand, smaller pores present stronger Im--Im interactions, which also stabilize the structure,\cite{Lewis2009} as mentioned above. For these reasons, these results cannot be directly compared with previous stability measurements based only in the relative crystal energies in vacuum. Furthermore, all these polymorphs are metastable with respect to the ZIF-zni dense polymorph, which implies that relative stability does not fully control polymorph selection. Within our model, the vacuum energies of the crystal structures follow the order: E$_{\rm{ZIF-zni}}$  $<$ E$_{\rm{ZIF-4}}$  $<$ E$_{\rm{ZIF-3}}$ $<$ E$_{\rm{ZIF-10}}$ $<$ E$_{\rm{ZIF-6}}$,\cite{Balestra2022} while \textit{ab initio} calculations show an inversion between the energies of ZIF-3 and ZIF-10.\cite{Lewis2009,Drholt2019} This trend does not coincide with the activation free energy trend.
The only experimental evidence available to the best of our knowledge is that ZIF-4 is the most commonly found polymorph within the range of synthesis conditions explored, while the other polymorphs are only obtained using very specific and controlled setups, as explained above.\cite{Park2006} 
This suggests that ZIF-4 is the most favoured structure in most of the cases, which is in agreement with all computational results, including those presented herein. The main synthesis feature that seems to control polymorph selection in Zn-Im-based ZIFs is the metal-to-ligand concentration ratio. We wish to assess whether different metal-to-ligand ratios lead to different PNC compositions, for example, an excess in ligand concentration could promote higher coordination of the Zn$^{2+}$ ions, and as a consequence, more ramified clusters. In previous work by Andarzi Gargari and Semino, the PNCs obtained when synthesizing Im$^-$ and 2-mIm$^-$ Zn-based MOFs were compared. The authors found that the PNCs formed with Im$^-$ were more ramified than those for 2-mIm$^-$.\cite{AndarziGargari2025} This makes the comparison between $\Delta$G$^{\transitionstatesymbol}$ values for different polymorphs not absolute, since the minimum that corresponds to the PNCs does not necessarily correspond to the same state in all the four polymorphs. In that way, a certain polymorph may have a lower $\Delta$G$^{\transitionstatesymbol}$ between PNC and transition states but the tagged PNC composition may not match what is typically obtained under the employed reactant concentration synthesis conditions. Moreover, we note that since the ZIF-10 intermediate-to-transition-state transformation happens at a later stage, the comparison between ZIF-10 and ZIF-4 is not straightforward.\\

\begin{figure*}[h!]
\centering
\includegraphics[width=0.4\textwidth]{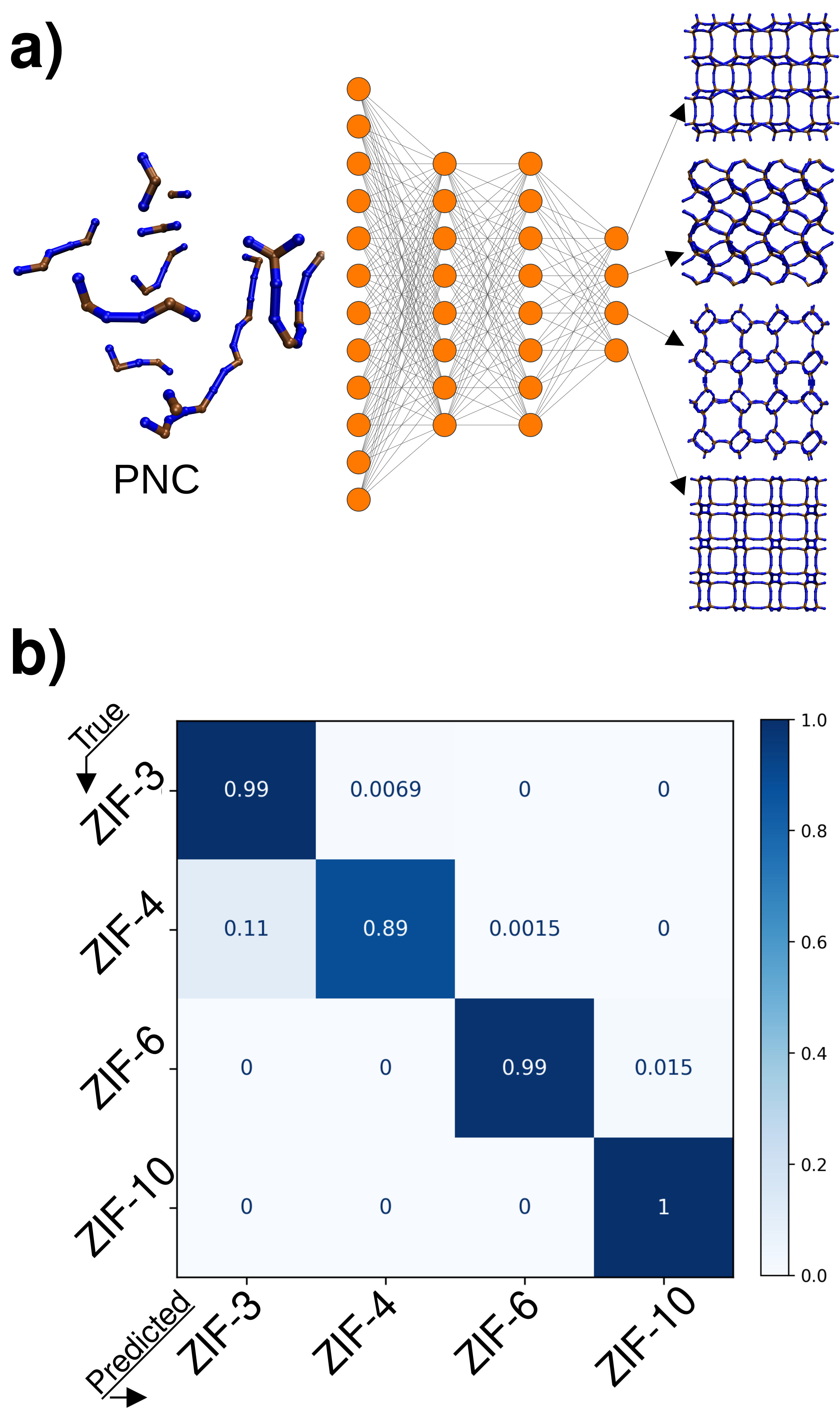}
\caption{\label{fig:confusion}{a) Scheme of the algorithm of the PNC classifier. The neural network architecture is reduced for allowing a better visualization. b) Confusion matrix of the classification algorithm. Results are normalized so that each row values sum one.}}
\end{figure*}

Evaluating whether each polymorph presents different PNC compositions or not is not trivial, since a wide variety of PNC microstates are compatible with the $q$=$q^{\rm{IS}}$ state for each polymorph. To tackle this challenge, we developed a classification algorithm as done above to see whether the PNCs can be differentiated or not. In this case, PNC global configurations are given as inputs instead of individual Zn-centred environments. The neural network outputs the probability that the input configuration is associated to each of the four polymorphs studied. Further details of this algorithm are provided in the Methods Section and in the SI. If the neural network is able to classify the PNCs according to their associated polymorph with high accuracy, it follows that they are distinguishable.   \textbf{Figure~\ref{fig:confusion}} b) shows the confusion matrix of the algorithm applied on the test set. The largest source of error arises from the misclassification of ZIF-4 PNCs as belonging to ZIF-3. This means that both PNCs present the highest degree of similarity. By looking at Figure \ref{fig:snaps} it can be seen that the PNCs that correspond to ZIF-3 and ZIF-4 typically have slightly lower chain lengths than those associated to the other two polymorphs.   Despite this error, the global accuracy of the algorithm in recognizing PNCs is around 97\%. This suggests that the PNCs that are generated along the synthesis of the different polymorphs are distinguishable, and thus, that polymorph selection may occur at the PNC stage of Zn-Im ZIFs self-assembly. This would also explain why metal-to-ligand ratio has a direct impact in the resulting synthesis product. 

\begin{figure*}[h!]
\centering
\includegraphics[width=0.7\textwidth]{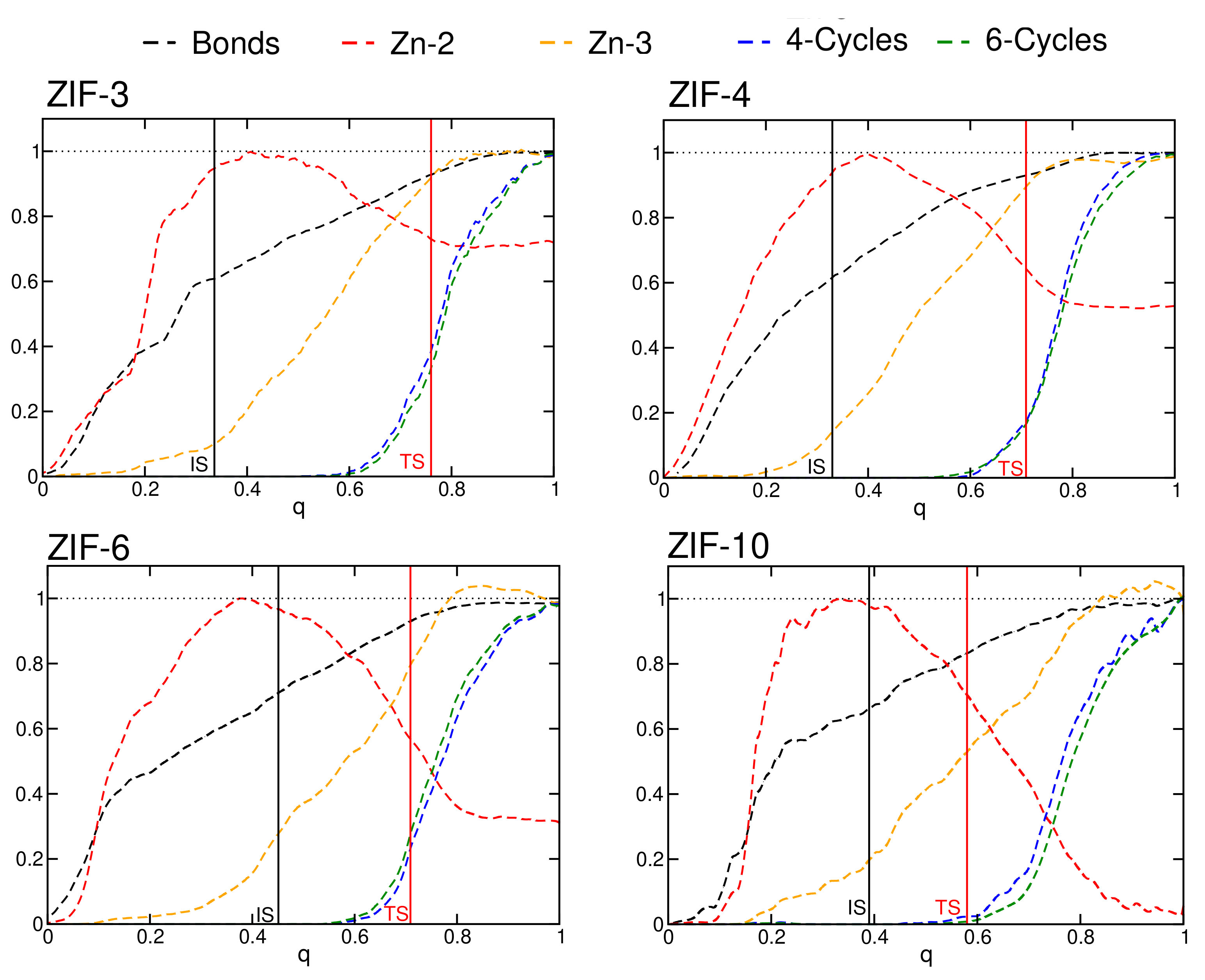}
\caption{\label{fig:features}{Average values of selected cluster features as a function of the reaction coordinate $q$ for each polymorph. The black curve corresponds to the number of Zn--N bonds, the red and orange curves to the number of 2 and 3-coordinated Zn ions respectively, and the blue and green curves to the number of 4 and 6-membered rings respectively. Each curve is normalized by the highest value of the feature along the whole path. The positions of the Intermediate and Transition states are signalled in each plot with black and red vertical lines.}}
\end{figure*}

\textbf{Figure \ref{fig:features}} shows a set of selected structural features as a function of the reaction coordinate $q$, to describe structural evolution in a more quantitative way. Zn--N bonds were plotted to measure the degree of connectivity as the reaction progresses. The amount of two and three-coordinated Zn$^{2+}$ ions (Zn-2 and Zn-3) and the amount of 4- and 6-membered rings comprised by Zn$^{2+}$ ions connected through Im$^-$ ligands, which together with 8-membered rings correspond to the only rings present in the final pores, are also included in the Figure. To facilitate the comparison, each plot was normalized by its maximum value along the reaction coordinate. As explained before, the configuration of the system is not completely defined by $q$. Instead, a manifold of the configurational space is compressed into each $q$ value. For that reason, the feature values should be averaged between those from all the configurations in the trajectories that share a specific target $q$. Surprisingly, the evolution of the selected features along the reaction path follows a similar pattern for all polymorphs. This means that regardless of the target structure, the optimal pore formation mechanism presents common features.

Bond formation seems to follow a three-steps mechanism: (i) a first rapid growth that corresponds to the early free ion combinations, which are highly reactive due to the effect of supersaturation, (ii) a second steady-state-like stage, in which the amount of bonds grows linearly with $q$, and (iii) a final stage where the bond number remains almost constant, which corresponds to a series of bond rearrangements without global connectivity changes. The location of the Intermediate State minimum of the free energy that corresponds to the PNC state is marked with a black vertical line in the plot, and in all cases it is formed within the steady-state stage of the bond formation process.\\

The evolution of the amount of Zn-2 and Zn-3 reveals information on the ramification of the formed clusters, since a cluster comprised only by 2-coordinated Zn$^{2+}$ ions is necessarily linear, while 3-coordinated ions are an indicator of a bifurcation. The curve that corresponds to the population of Zn-2 presents a maximum close to the PNC state and then it starts decreasing as the Zn-3 curve starts to grow. This point can be associated with the formation of ramified clusters. By looking at the position of the Intermediate State we can observe that the PNCs are mostly linear chains, with a small amount of Zn-3, as can be also seen qualitatively in the snapshots shown in Figure~\ref{fig:snaps}. The relative amount of Zn-2 and Zn-3 in the final pore structures differs between polymorphs, with ZIF-10 containing Zn-3 only and ZIF-3 having the highest associated Zn-2 population.

The amount of $n$-membered rings present in the formed clusters serves as an indicator of the emergence of closed pore-like structures. It is worth to mention that 3- and 5-membered rings were observed multiple times along the trajectories, despite not being part of the final pores. This is a positive sign since it is known that the amorphous-like aggregates that follow PNC formation present these kinds of odd $n$-membered rings.\cite{AndarziGargari2025,Balestra2022} 

\section{Conclusions}

In this work, molecular simulations combined with machine learning techniques were employed to investigate the self-assembly mechanism of Zn(imidazolate)$_2$ metal–organic frameworks and to determine at which stage of the synthesis mechanism is the final polymorph selected. Using adaptive path collective variable metadynamics, representative ensembles of configurations corresponding to both amorphous intermediates and pre-nucleation clusters were generated for several polymorphs.

The thermodynamic analysis of amorphous-to-crystal transitions indicates that the relative stability between amorphous and crystalline phases depends on the polymorph and it is significantly influenced by solvent effects. Only ZIF-4 and ZIF-10 presented a negative free energy difference between solvated amorphous and crystalline forms at the thermodynamic conditions studied. Crystalline porous phases are capable of adsorbing more solvent molecules than the more compact amorphous phases, but the presence of reactive sites in the amorphous phases strengthens solvent-framework interactions. 
Thus, the presence of the solvent can contribute to the relative stabilization of either of the two phases. Polymorphs with larger pore sizes present a net stabilization of the crystalline phase while the opposite occurs for polymorphs with smaller pores.
Structural characterization and neural network classification show that amorphous intermediates associated with different polymorphs exhibit distinguishable structural features, suggesting that these phases are not structurally equivalent among polymorphs.

A similar approach applied to pre-nucleation clusters reveals that these early-stage species can also be differentiated according to the polymorph to which they are associated. The analysis of free energy profiles for pore formation highlights a multi-step mechanism involving the formation of oligomeric clusters followed by their growth and structural rearrangement into ordered frameworks. Despite similarities in the overall pore formation mechanism between polymorphs, quantitative differences in the free energies plots are observed. Moreover, neural network classifiers confirm that the pre-nucleation clusters are structurally different for the four polymorphs investigated.

These results suggest that structural differences between polymorphs emerge at early stages of the self-assembly process, already at the level of pre-nucleation clusters formation, and persist through the formation of amorphous intermediates and up to the crystalline phase.

The methodologies and tools used in this work can be adapted to model the self-assembly process of any other metal-organic framework, provided that an appropriate reactive force field is available. We hope that this work will inspire further research on simulation and machine learning-assisted self-assembly studies.

\section{Methods}

\subsection{Simulation setup}

All the simulations were performed using the LAMMPS software\cite{lammps} coupled with the PLUMED package.\cite{Tribello2014} 
nb-ZIF-FF\cite{Balestra2022} was used to model all ZIFs. nb-ZIF-FF adequately reproduces the properties of ZIF-4 and several of its polymorphs, including those that result from its thermal amorphization and melting,\cite{Mendez2024} its high pressure phases\cite{Mendez2024_2} and ZIF-3, ZIF-6 and ZIF-10 crystals.\cite{Balestra2022} 

All results were obtained from molecular dynamics simulations made in the NPT ensemble at $T=400$ K and $P=1$ bar, which reproduces the experimental solvothermal synthesis conditions of ZIF-4.\cite{Park2006} The timestep was set to 0.25 fs. 
The amorphous-to-crystal transitions systems comprised a total amount of 128 Zn$^{2+}$ and 256 ligand ions. This corresponds to a 2x2x2 supercell for ZIF-4, ZIF-3 and ZIF-6, and a 2x1x2 supercell for ZIF-10. Thus, the system volumes lie between 27,000 and 55,000 \AA$^3$ depending on the polymorph.
The ions-to-pore systems comprised a total amount of 60 Zn$^{2+}$, 120 ligand ions and 583 DMF solvent molecules, which gives a total of 4998 atoms including dummy atoms. The box size was around $\sim$50x50x50 \AA$^3$. 

The C$\mu$MD scheme developed by Karmakar \textit{et al.}\cite{Perego2015,Karmakar2019} in spherical coordinates was used to keep the concentration of Zn$^{2+}$ and ligand ions constant along the pore formation simulations. This technique was successfully applied to study the nucleation sodium chloride\cite{Karmakar2019} and a similar scheme was used for studying the growth mechanism of ZIF-8 at constant reactant concentration.\cite{AndarziGargari2026} The details of the procedure are briefly explained in the SI.

\subsection{Adaptative Path collective variables}

MOF self-assembly consists of a collection of rare events that involve the formation and breaking of coordination bonds, solvent displacements and entropy reduction. These reactions take place at timescales in the order of hours experimentally (see Table S2 in the SI), which means that they are highly activated.\cite{Han2022}
On the other hand, atomistic simulations are limited to the $\mu$s timescale at the most. For this reason, it is mandatory to implement an enhanced sampling technique to sample the reaction pathway in a reasonable amount of computational time.
Metadynamics and its related techniques are among the most used algorithms to efficiently sample rare events.\cite{Laio2002}
In metadynamics, a CV, which is a function of the atomic coordinates, serves as reaction coordinate.
The metadynamics algorithm will dynamically add a bias potential to the system in order to force it to explore the full range of possible CV values.  
Selecting the CV is tricky: not only it must distinguish reactant and product states but also it must be able to drive the system from reactants to products and vice versa when a bias potential is imposed on it.
Typical CVs used in self-assembly simulations include coordination numbers, Steinhardt parameters, environmental similarity and distance from a reference configuration.\cite{Balestra2022,Karmakar2021,Neha2022}
None of these CVs allowed to reversibly probe the pore and amorphous intermediate formation reactions studied in this work, neither did the combinations of pairs or triplets of them. The main reason for this is that Zn--ligand bonds are stronger ($\sim10k_bT$\cite{Mendez2025}) than other types of softer interactions usually found in crystallization processes. This makes the reorganization of a defective structure much more time-consuming, and as a consequence, it leads the system to get trapped in local energy minima intermediate states. To overcome this problem we employed a path collective variable approach.\cite{DazLeines2012} Within this technique, a large set of CVs $\vec{\xi}$ can be used simultaneously to model the transition. Using this whole set of CVs to bias the trajectory in a straightforward manner would not be possible, as it would prevent convergence. Instead, a curve is constructed in the high dimensional CV space (the \emph{path}) and the degree of advance within this curve is used as a single CV coordinate ($q$) that contains the information of the full set of CVs. 
Another variable $z$ will be defined to measure the minimum distance in the CV space between the curve and the CV values of the system. In order to maintain the system close to the target curve, $z$ is constrained to have small fluctuations by the addition of a harmonic bias potential $U(z)$ of the type:
\begin{equation}
    U(z) = \frac{k_z}{2} z^2
\end{equation}
Where $k_z$ is a force constant. In this way, $z$ is not constrained to be exactly zero, and the system can explore the vicinity of the curve. 
The initial guess of the curve itself will be generated by monitoring the values of all CVs along a trial trajectory in which the reaction takes place. The path will then be updated towards the optimal transition path by following the procedure developed by G. Díaz Leines \textit{et al}.\cite{DazLeines2012}
It consists of performing a metadynamics simulation in which $q$ serves as CV, so that several transitions between reactants and products will be sampled. Since the system has a certain freedom to deviate from the defined curve, it will tend to approach more favourable pathways. The curve is then updated according to the following equation:
\begin{equation}
    \vec{s}(q)=\langle \vec{\xi}\rangle_q
\end{equation}
Where $\vec{s}(q)$ represents the curve in CV space and $\langle \vec{\xi}\rangle_q$ is the mean value of the CVs among all the configurations of the trajectory that share the same value of $q$. During this procedure, the $N$ points that describe the path are recomputed in each update step in order to remain equidistant. 
The protocol for generating the initial guess trajectories varies according to the system considered and is detailed in the SI. This trial trajectory does not need to correspond to the optimum pathway since it will be further refined. A set of $N=20$ equidistant points in CV space was used to describe the curve numerically. 
The path update was executed each 10$^6$ simulation steps. The path update procedure is monitored by measuring the distance between consecutive paths. The convergence criterium is reached when this difference becomes lower than a threshold value of 0.05 (see Figure S3 of the SI). The effect of the update procedure in the path is depicted in Figure S4 of the SI, where we show, as an example, the evolution of the Zn--N coordination number along the path as the update takes place.\\ 
Once the path is converged, a final production metadynamics run is performed keeping the path fixed. All the results shown in the Results Section come from these simulations. The free energy error calculation procedure is detailed in the SI.\\
For the implementation of the adaptive path collective variable scheme, we used the PATHCV Plumed package developed by the group of Bernd Ensing, which can be found in \url{https://github.com/Ensing-Laboratory/PathCV}.\cite{DazLeines2012}

The set of CVs used for defining the path comprises a combination of a small amount of global features and large amount of local features. 
The global features are: number of Zn--ligand and Zn--solvent bonds, the amount of Zn$^{2+}$ and Im$^-$ ions in the reaction region in the case of pore formation; and number of Zn--ligand bonds and the volume in the case of the amorphous-to-crystal simulations. Inspired by environment similarity descriptors commonly employed for describing nucleation,\cite{Piaggi2019} we also included local features that consider the degree of similarity between the configuration of the system and a template of the target structure.
Each of these CVs $\xi_i$ will measure the presence of a specific atom type in the location of a template site $i$. This is done in practice by placing a virtual atom in the simulation box at the position of the template site $\vec{r_i}$ and computing the overlap between a Gaussian centred in the virtual atom and all the species of the tagged type:
\begin{equation}
    \xi_i = \sum_j e^{-\frac{(\vec{r_i}-\vec{r_j})^2}{2\sigma^2}}
\end{equation}
where the index $j$ covers all the atoms in the simulation box of the tagged type (Zn, N, C), $\vec{r_i}$ and $\vec{r_j}$ are the positions of the virtual atom $i$ and the atom $j$ respectively, and $\sigma$ is a tunable parameter. In this way, each of these CVs can be seen as an on/off function that will output zero or one according to whether an atom is located at the template site or not. A total number between 132 and 372 CVs were used for the pore formation simulations, depending on the size of the pore, and a total of 514 CVs were used for each amorphous-to-crystal simulation. It is important to highlight that all the CVs used are permutation invariant, so no tagged Zn$^{2+}$ or Im$^-$ is forced to occupy a predefined place in the self-assembled structure. This is essential to account properly for the entropic changes along the reaction pathways. 

\subsection{Classification algorithms}

To classify local Zn-centred environments into polymorphs, we implemented neural network classifiers. These algorithms require that chemical environments are described through a set of features that will serve as inputs. For doing so we used Behler-Parrinello Symmetry Functions\cite{Behler2007} (BPSF), which are translation, rotation, and permutation invariant descriptors that were initially developed for the construction of neural network potential energy surfaces and were also recently used for environment classification tasks.\cite{Mendez2024,https://doi.org/10.48550/arxiv.2604.09084} In this way, each Zn-centred environment is described by a set of 24 BPSFs that only depend on the positions of the neighbour Zn and N atoms within a radius of 13 \AA. Information from all other atom types was neglected. 
A description of the implemented BPSF functional form and parameters is given in the SI. The neural network performs non linear combinations of these features to obtain as a result four outputs, each one of them representing the probability that the input comes from a simulation of a specific polymorph. 

For the amorphous phases classification, we constructed a dataset of $\sim$ 320,000 BPSF environments labelled according to the polymorph from which they were produced. For the PNC classification, 120,000 environments were used. These data were obtained from the metadynamics simulations that were used to perform all the previous analyses. For each polymorph, only the configurations that correspond to $q$-values of $q^{\rm{IS}}\pm$0.025 were considered as valid amorphous or PNC states. In the case of PNC analysis, only Zn$^{2+}$ ions that are inside the growth region of the simulation box were considered, since the ions that lie outside this region are forced to remain zero coordinated.

The data set was split into train and test sets in 80:20 proportion. The train set was used to fit a fully connected feed forward neural network with 2 hidden layers of 64 nodes each. The output layer consists in 4 nodes that pass through a softmax function that ensures that they sum one and are all positive. Each node value corresponds to the probability of that input being associated to the synthesis pathway of one of the four polymorphs.

In the case of PNCs, we performed a cluster composition-based classification, and not just predictions at a local Zn$^{2+}$ level. To do so, the global prediction of a configuration was obtained from the combination of the local Zn$^{2+}$ environment predictions according to the following procedure:
(i) For each Zn$^{2+}$ ion $i$ in the configuration, the BPSF descriptors are computed, and the neural network outputs the probabilities $P_{ij}$ that the ion $i$ comes from a simulation corresponding to polymorph $j$.
(ii) Based on a Naive Bayes criterium, the global prediction corresponds to the polymorph $j$ that maximizes the log-likehood of the individual results $\mathcal{L}_j$:\cite{Sayed2022}
\begin{equation}
    \mathcal{L}_j = \sum_iln(P_{ij})
\end{equation}

\begin{acknowledgement}
This work was funded by the European Union ERC Starting grant MAGNIFY, grant number 101042514. This work was granted access to the HPC resources of CINES under the allocations A0170915688 and A0190915688 made by GENCI. 
\end{acknowledgement}


\newpage
\setcounter{figure}{0}
\setcounter{table}{0}

\renewcommand{\thetable}{S\arabic{table}}  
\renewcommand{\thefigure}{S\arabic{figure}} 
\section{Supporting Information for:\\
Machine Learning and Molecular Simulations Reveal Mechanisms of ZIFs Polymorph Selection}

\tableofcontents

\addcontentsline{toc}{section}{Simulation details}
\section{Simulation details}

Within the nb-ZIF-FF forcefield,\cite{Balestra2022} Zn--N interactions are represented by Morse potentials that allow bond formation and breaking. This force field also incorporates dummy atoms attached to the Zn and N atoms to correctly reproduce the angular distribution of ligands around a Zn$^{2+}$ ion.
For the N,N-dimethylformamide (DMF) solvent, a flexible version of the CS2 potential developed by Chalaris and Samios was employed.\cite{Chalaris2000} Intramolecular bond, angular and dihedral parameters were taken from the CHARMM general force field.\cite{Vanommeslaeghe2009}
Both nb-ZIF-FF and CS2 parameters were re-adjusted to reproduce the bond energies between Zn--N(ligand) and Zn--O(DMF) computed by Density Functional Theory calculations, as done in our previous works, where all the force field parameters can be found.\cite{Mendez2025} 
Nosé-Hoover thermostats and barostats were used with damping times of 100x and 1000x the timestep respectively in all cases. Periodic boundary conditions were always employed in all directions. Long range electrostatics were treated with the particle-particle particle-mesh algorithm, using a relative error threshold for the forces of 10$^{-6}$.

\addcontentsline{toc}{section}{Constant chemical potential simulations}
\subsection{Constant chemical potential simulations}

\begin{figure}[H]
\centering
\includegraphics[width=0.45\textwidth]{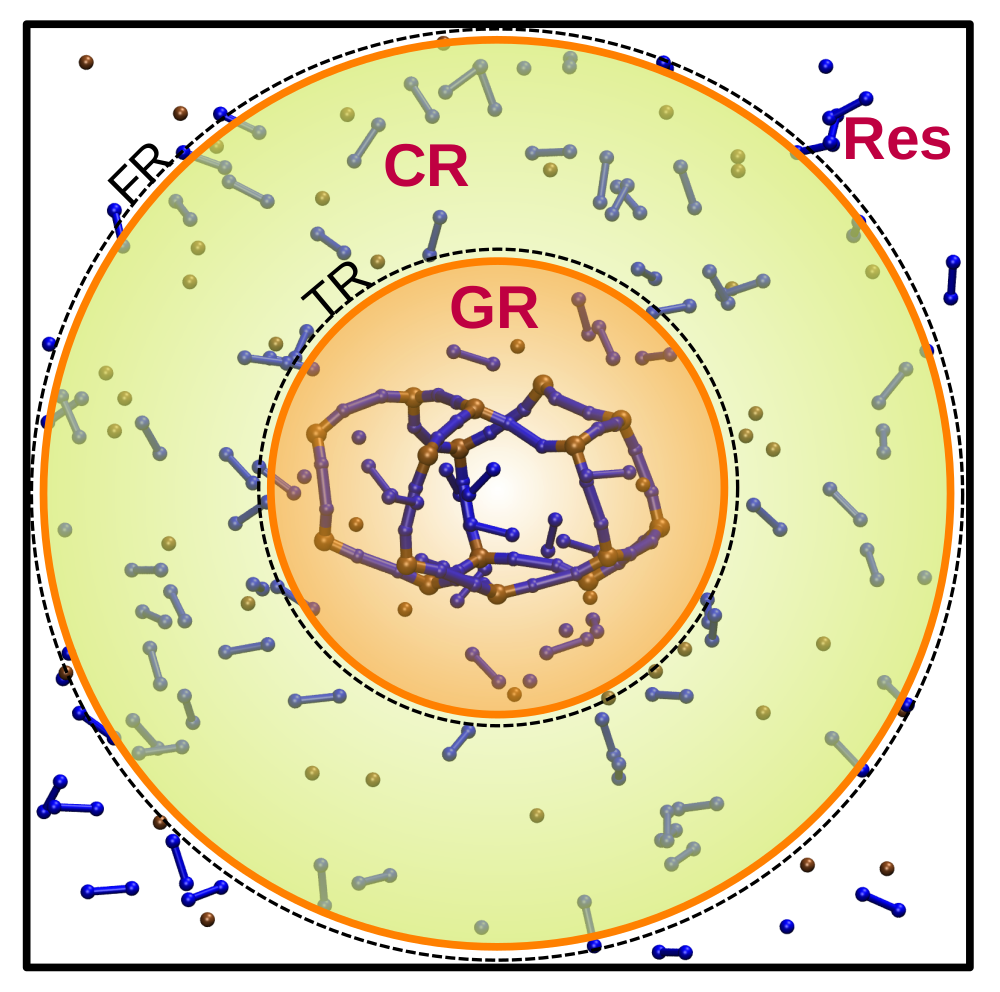}
\caption{\label{fig:cmumd}{C$\mu$MD scheme for the ZIF-4 pore formation. CR: Control region, where [Zn$^{2+}$] and [Im$^-$] are kept fixed. GR: Growth region, where the self-assembly will take place. Res: Reservoir. TR: Transition region. FR: Force region. Solvent molecules are not displayed. Zn$^{2+}$ ions are represented as ochre spheres and ligand moieties by blue sticks located at the position of the N atoms.}}
\end{figure}

To set up the constant chemical potential molecular dynamics simulations (C$\mu$MD) to model pore formation, we followed the procedure described below. First, the simulation box is divided in three regions (see Figure \ref{fig:cmumd}): (i) the growth region (GR), where the pore formation will take place, represented by a sphere centred in the middle of the box, (ii) a control region (CR) where Zn$^{2+}$ and Im$^-$ concentrations will be kept constant, which corresponds to the region between two spherical caskets of radius $r_{in}$ and $r_{out}$ and (iii) a reservoir region (Res) that comprises the rest of the box and will serve as buffer for the remaining ionic species. Within this setup, the reservoir region comprises approximately half of the simulation box.
This technique is available within the Plumed package,\cite{Tribello2014} and was obtained from the repository of Tarak Karmakar: \url{https://github.com/Tarakk/plumed-cumd}.\cite{Karmakar2019,Perego2015}

Within this method, the solute concentration ($C$) in the CR is calculated as
\begin{equation}
    C=\frac{1}{V^{CR}}\sum_{j=1}^Nf(r_j)
\end{equation}
where $V^{CR}$ is the CR volume and $r_j$ is the distance of a $j$th
particle from the box center. The $f(r_j)$ is a continuous and
differentiable switching function that counts atoms belonging to
the CR, defined by:
\begin{equation}
    f(r_j)=\frac{1}{1+e^{-(r_j-r_{in})/\alpha}}\frac{1}{1+e^{(r_j-r_{out})/\alpha}}
\end{equation}
where $\alpha$ is a parameter that controls the switching functions steepness. The function $f(r_j)$ has a value of 1 when the solute is inside the CR and 0 when it is outside.

The force that restraints the instantaneous solution concentration $C$ at a target value ($C_0$) has the following expression:
\begin{equation}
    F(r) = \kappa(C-C_0)G(r)
\end{equation}
where $\kappa$ is the force constant. $G(r)$ is a bell-shaped function that localizes the force $F(r)$ at $r_F=r_{out}+w_F$. This function is defined as:

\begin{equation}
    G(r)=\frac{1}{2\sigma}\frac{1}{1+\mathrm{cosh}((r-r_F)/\sigma))}
\end{equation}
where $\sigma$ is a broadening parameter.

The C$\mu$MD parameters are listed in Table \ref{cmumdtable}. Their choice was based on previous setups.\cite{Karmakar2019} 
The selected concentration of reactants was higher than in typical synthesis experiments, where values of $\sim$10$^{-6}$-10$^{-2}$ M are used.\cite{Park2006} The reason behind this choice is that such low concentrations would require much larger simulation boxes in order to have a significant amount of reactant species in the Control Region.  

\begin{table}[H]
\small
  \caption{C$\mu$MD parameters. C(Zn$^{2+}$) and C(Im$^{-}$) correspond to metal and ligand target concentrations respectively. $\kappa$ corresponds to the force constant of the constraint. $r_{in}$  and $r_{out}$  correspond to the inner and outer radii of the Control Region. $w_{F}$ corresponds to the width of the force region.}
  \label{cmumdtable}
  \begin{tabular*}{0.95\textwidth}{@{\extracolsep{\fill}}llllllll}
    \hline
     $C_0$(Zn$^{2+}$) & $C_0$(Im$^{-}$) & $\kappa$ & $r_{in}$  & $r_{out}$   &  $w_{F}$ &$\alpha$ & $\sigma$  \\
    \hline
0.86 M     & 1.72 M    &  10000 kJ/mol/nm$^3$ & 1.7 nm & 2.5 nm & 0.2 nm & 0.05 nm & 0.05nm \\
    \hline
  \end{tabular*}
\end{table}

\begin{figure}[H]
\centering
\includegraphics[width=0.55\textwidth]{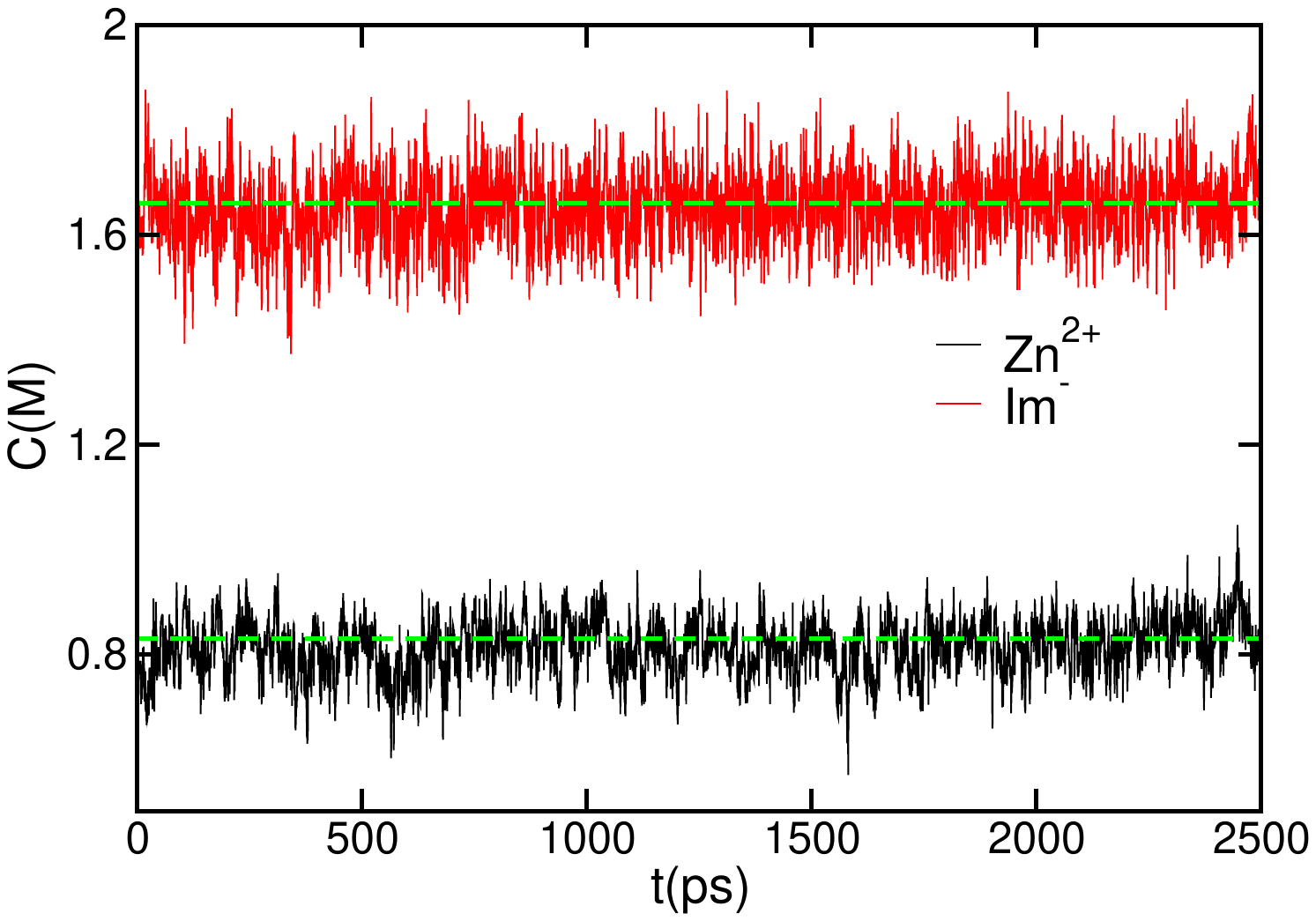}
\caption{\label{fig:cmumd_conc}{Zn$^{2+}$ and Im$^-$ concentrations during a typical C$\mu$MD simulation in the control region. The simulation corresponds to the initial guess trajectory of a ZIF-4 ions-to-pore setup. Green lines indicates the target value imposed by the constraint.}}
\end{figure}

For validation, we plotted the time-evolution of the concentrations of reactants for a typical simulation where a ZIF-4 pore decomposes to its non-connected ions in Figure \ref{fig:cmumd_conc}. It can be observed that the concentrations fluctuate around the target values at all times.

\addcontentsline{toc}{section}{Pore system preparation}
\section{Pore system preparation}

Pore structures are prepared as follows: 
(i) crystal configurations displayed in Figure 1 are obtained from the experimental \emph{cif} files\cite{Park2006} and equilibrated using our force field.\cite{Balestra2022}
(ii) A pore centre is identified and the Zn$^{2+}$ and ligand moieties that comprise the pore are selected to build the template.
(iii) The pore template is placed in the centre of an empty box of 50 x 50 x 50 \AA$^3$. Extra Zn$^{2+}$ and ligand species are randomly added in the empty region of the box until a total of 60 Zn$^{2+}$ and 120 ligands is reached.
(iv) DMF solvent molecules are added in the box via a Grand Canonical Monte Carlo algorithm until 583 solvent molecules are incorporated.\cite{Metropolis1949} The system is then equilibrated under NPT conditions until the equilibrium density is reached. The central template is kept rigid during this stage to prevent its decomposition. 
(v) The C$\mu$MD\cite{Karmakar2019} scheme is activated to keep the Zn$^{2+}$ and ligand ions concentration constant at values of 0.83 M and 1.72 M respectively in the control region (see Figure \ref{fig:cmumd}). The connectivity between Zn$^{2+}$ and ligands was constrained to be zero outside the reaction region to prevent aggregation in the control region or in the reservoir. The central cluster was allowed to move but its centre of mass position and orientation were kept constant by constraining to zero its total linear and angular momenta.

\addcontentsline{toc}{section}{Generation of the initial path collective variable guess trajectories}
\section{Generation of the initial path collective variable guess trajectories}

For the amorphous-to-crystal transitions, crystal configurations were melt-quenched by imposing a set of temperature ramps. First, the system was equilibrated during 500,000 simulation steps at 400 K. Then, the temperature was linearly increased up to 1000 K during 500,000 steps. Subsequently, the system, in liquid state, was equilibrated at 1000 K during 2,000,000 steps and further cooled down to 400 K during 1,000,000 steps. Finally, the obtained amorphous structure was equilibrated at 400 K  during 500,000 steps.

For the pore systems, the formed pore is selected as initial state and a harmonic constraint is imposed to regulate the number of Zn--N bonds. This constraint will smoothly force the cluster to decrease its connectivity from the initial value to zero along 2x10$^6$ time steps. The final structure corresponds to a fully disconnected system, which corresponds to what we call the reactant state in the article text.
Within this procedure, the concentration of Zn$^{2+}$ and ligand ions in the control region is kept constant at values of 0.86 and 1.72 M respectively. The total linear and angular momentum of the central cluster are kept fixed at zero to avoid random rotations and displacements. 

\addcontentsline{toc}{section}{Path update convergence}
\section{Path update convergence}

The individual collective variables (CVs) that compose the path were normalized in such a way that in the initial guess trajectory the maximum and minimum values correspond to 0 and $N_{max}$, were $N_{max}$ was set to 1 for local CVs and 5 for global CVs. In this way, a higher relative importance is given to global CVs in the calculation of the path than local ones.
The final path variables $q$ and $z$ are then dimensionless scalar quantities.
The force constant applied to $z$ was set to 4000 and 5000 kJ/mol for the pore formation and amorphous crystallization simulations respectively.

\begin{figure}[H]
\centering
\includegraphics[width=0.55\textwidth]{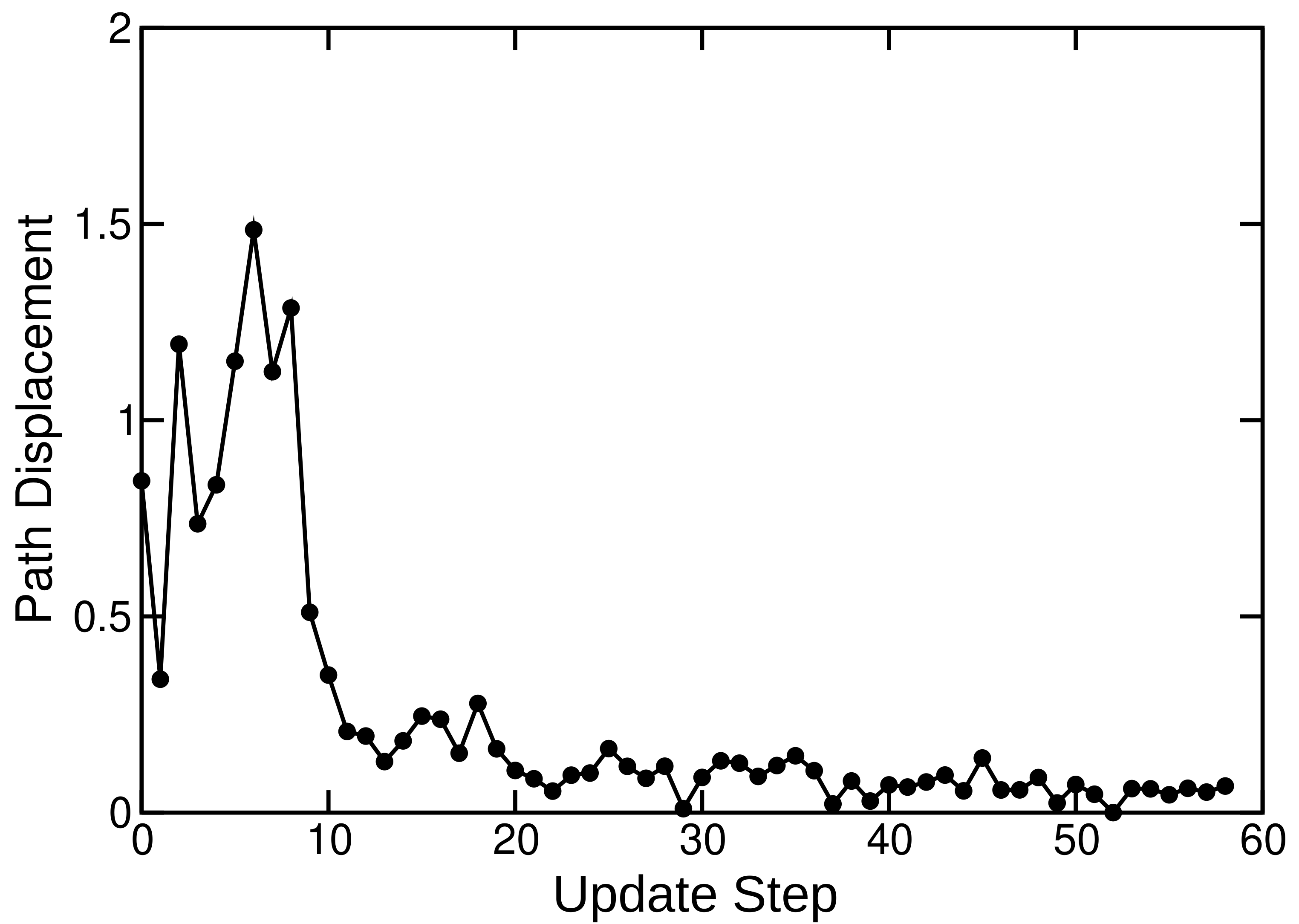}
\caption{\label{fig:update}{Path displacement with respect to the previous configuration as a function of update step. The data shown corresponds to the path update of the ZIF-4 amorphous-to-crystal transformation.}}
\end{figure}

\begin{figure}[H]
\centering
\includegraphics[width=0.55\textwidth]{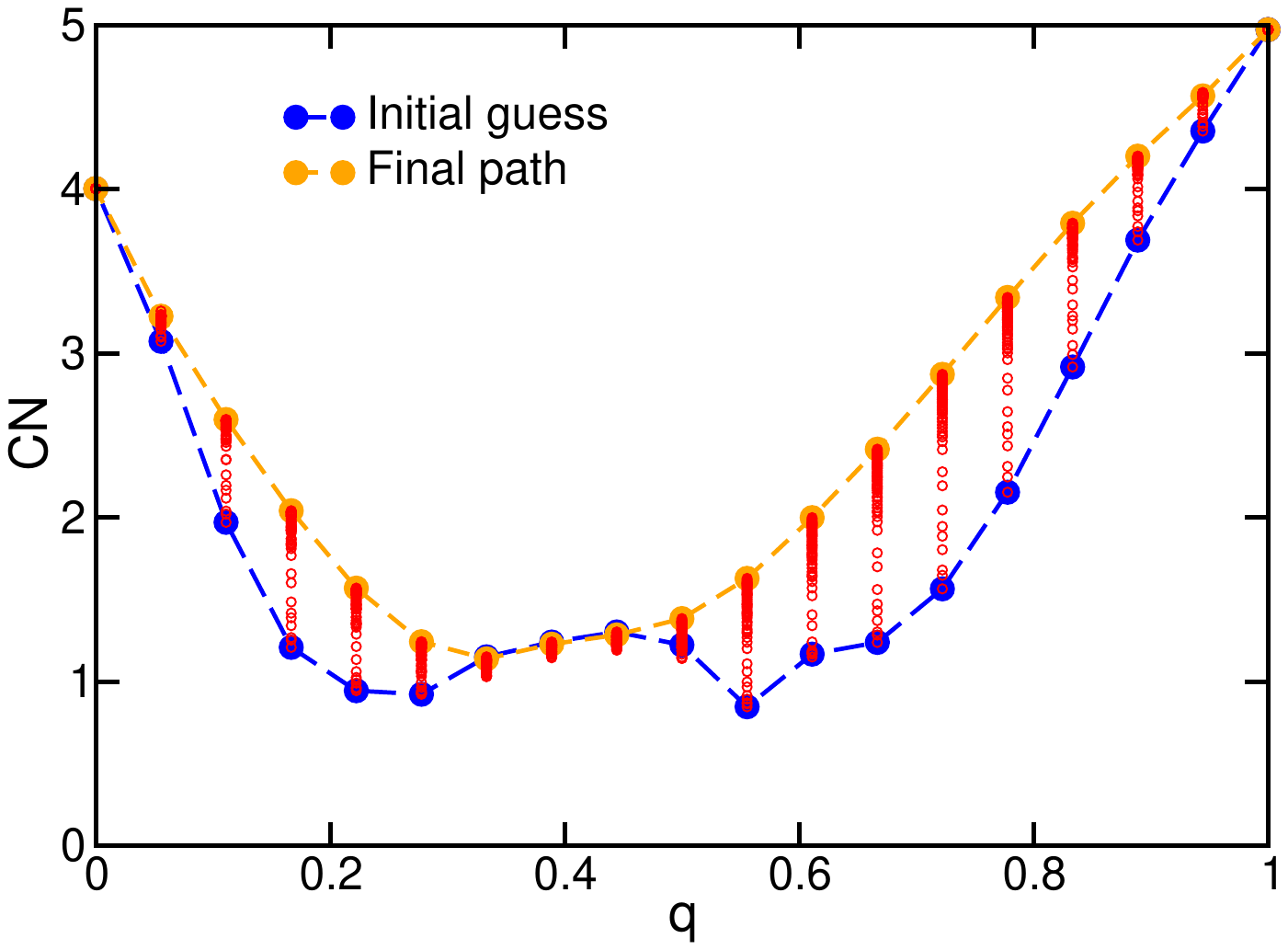}
\caption{\label{fig:update_path}{Evolution of the Zn--N coordination number (CN) along the path during an optimization process. The data corresponds to the same transformation as in Figure \ref{fig:update}.}}
\end{figure}

\addcontentsline{toc}{section}{Metadynamics setup}
\section{Metadynamics setup}

The parameters for the metadynamics\cite{Laio2002} simulations were set as follows: the initial Gaussian height was set equal to $kT$ ($T=$ 400 K), the Gaussian widths were 0.5
the pace for Gaussian deposition was equal to 75 fs (300 steps).
Ten parallel walkers were employed to accelerate the convergence of the amorphous-to-crystal simulations and five in the case of the pore formation runs.\cite{Raiteri2005} Each walker evolves independently from the others, but they all share the same bias potential obtained from the addition of Gaussian terms.
The total time for each simulation, comprising all the walkers, was around 25 ns. 
Upper and lower boundaries to the variable $q$ were included to prevent the system from exploring non-physical regions. This is done by applying harmonic constraints when the system goes above $q=1.05$ and below $q=-0.05$. The force constant used for this constraint was set to 10000 kJ/mol.

\addcontentsline{toc}{section}{Metadynamics convergence and uncertainty calculation}
\section{Metadynamics convergence and uncertainty calculation}

\begin{figure}[H]
\centering
\includegraphics[width=0.55\textwidth]{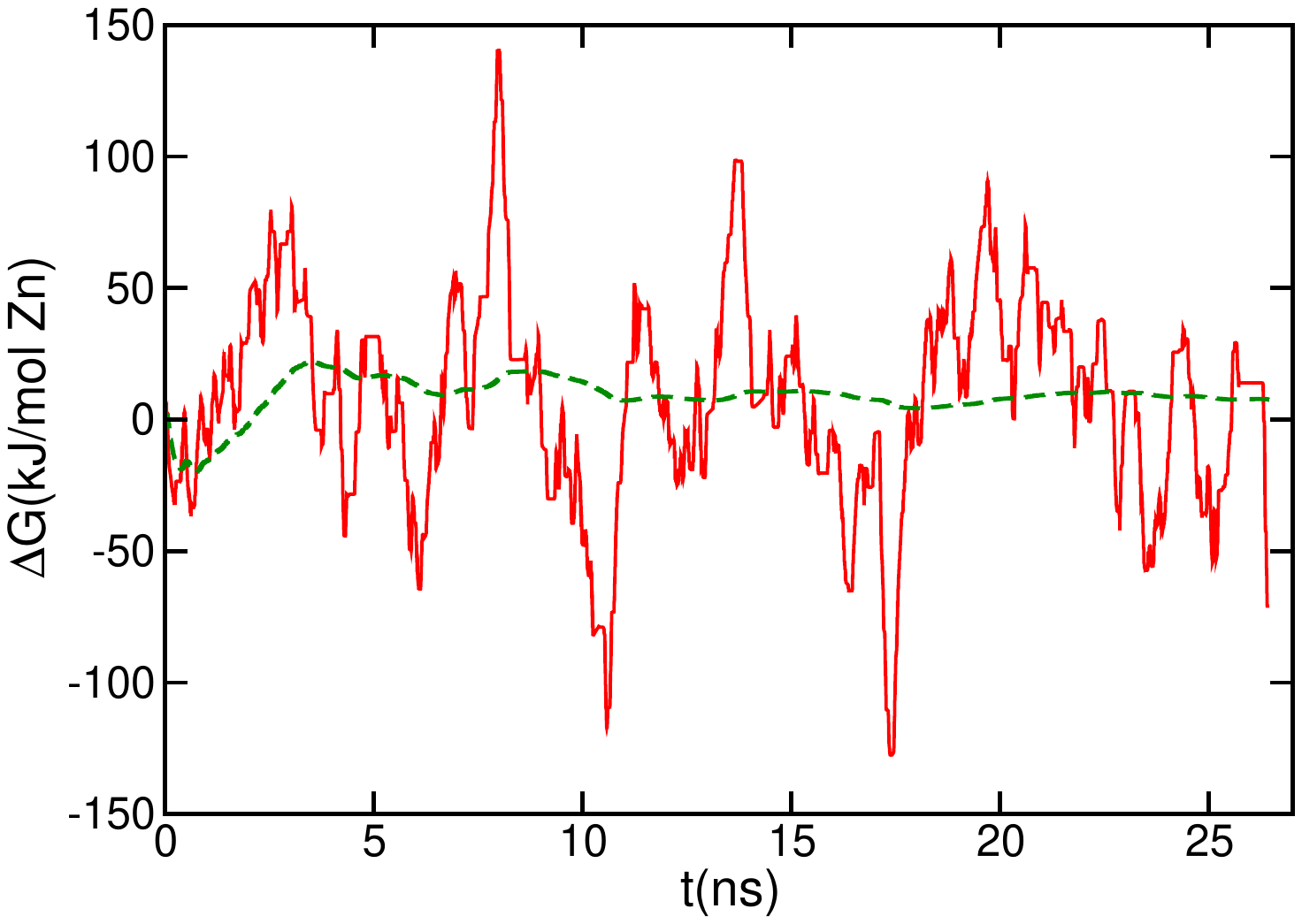}
\caption{\label{fig:estimator}{Free energy difference between the two minima as a function of simulation time for ZIF-3 pore formation (red curve). The time average of the result is shown in green.}}
\end{figure}

To analyse the convergence of the metadynamics simulations we saved the negative bias potential at different times during the simulations.
The results were aligned at the global minimum and the average and standard deviation were obtained as a function of the CV value. 
To check the convergence, we plotted the difference between the two lowest minima as a function of time. Figure \ref{fig:estimator} shows the corresponding plot for the ZIF-3 pore formation as an example. The plots for all the other reactions studied are qualitatively similar.

\begin{figure}[H]
\centering
\includegraphics[width=0.55\textwidth]{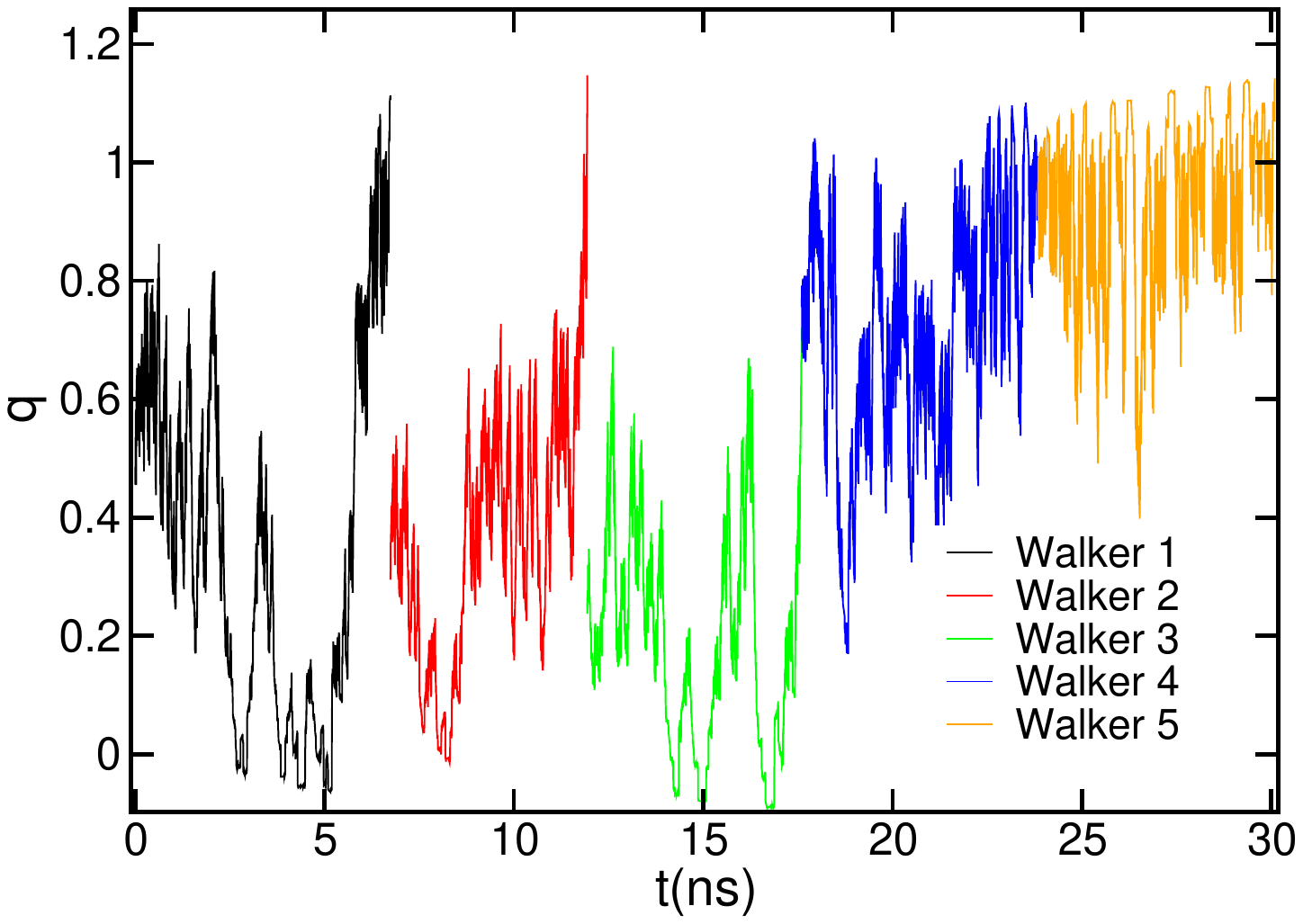}
\caption{\label{fig:walkers}{Collective variable $q$ as a function of time for each walker. For clarity purposes the plots were displaced in time to not overlap, in practice the five walkers are run simultaneously.}}
\end{figure}

We also plotted the CV $q$ as a function of time for each walker in Figure \ref{fig:walkers} to check that all the relevant $q$ values were visited multiple times.

\begin{figure}[H]
\centering
\includegraphics[width=0.55\textwidth]{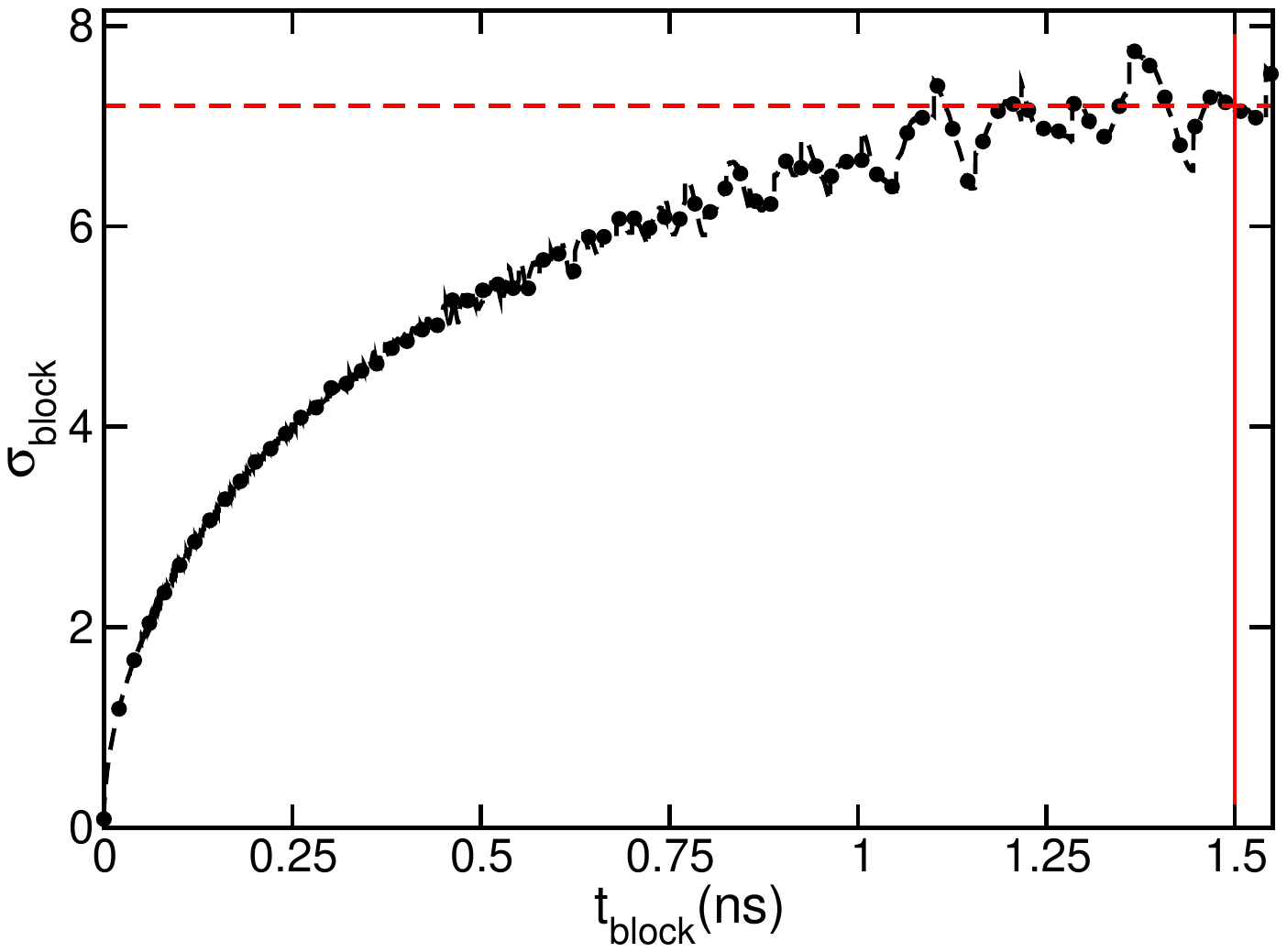}
\caption{\label{fig:blocks}{
Standard deviation of the free energy as a function of block size, for the same reaction as in Figure \ref{fig:estimator}. In red lines mark the value considered as optimal block size and its corresponding standard deviation.}}
\end{figure}

We employed the block averaging technique developed by Bussi and Tribello\cite{Bussi2019} to compute the error.
To estimate the optimal block size for which the data is uncorrelated, we computed the standard deviation of the free energy as a function of block size. 
Results are shown in Figure \ref{fig:blocks}. When the individual block values become uncorrelated, the standard deviation reaches a plateau. According to this criterion, we averaged data from blocks of 1.5 ns. 
This procedure was performed for all the reactions studied. The resulting plots are qualitatively similar to the one shown in Figure \ref{fig:blocks}.
Finally, the standard deviation was divided by $\sqrt{N_{blocks}}$ where $N_{blocks}$ is the number of statistically independent blocks of data obtained by dividing the total simulation time by the block size.

\addcontentsline{toc}{section}{Transformations of the collective variable space}
\section{Transformations of the collective variable space}

To transform the free energy associated to the CV $q$ to that associated to the Zn--N coordination number ($CN$) a reweighing procedure was employed.\cite{Schafer2020-sd}

Free energy $G(CN)$ can be obtained from the Boltzmann inversion of the probability $\mathcal{P}(CN)$:

\begin{equation} \label{gcn}
  G(CN)   = -k_bT ln(\mathcal{P}(CN))
\end{equation}

Where $k_b$ is the Boltzmann constant and $T$ is the temperature. Then, $\mathcal{P}(CN)$ can be obtained from a biased simulation where a bias potential $U^{bias}(q)$ is applied by performing a weighted histogram:

\begin{equation} \label{prob}
    \mathcal{P}(CN) = C\langle\delta(CN-CN(\vec{r})) e^{\beta U^{bias}(q)}\rangle_{biased}
\end{equation}

\noindent
where $C$ is an irrelevant constant.

\addcontentsline{toc}{section}{Free energy of solvent adsorption via hybrid MC/MD in the osmotic ensemble}
\section{Free energy of solvent adsorption via hybrid MC/MD in the osmotic ensemble}

We performed hybrid Monte Carlo(MC)/MD simulations to compute the adsorption free energy of crystal and amorphous structures. In these simulations, solvent molecules can be added or removed from the simulation box by the MC algorithm while the configuration and volume of the system relax at constant $T$ and $P$ during the MD runs.\cite{Rogge2019_2,Jeffroy2008} The thermodynamic ensemble sampled thus corresponds to the osmotic ensemble in which $N_{host}$, $P$, $T$ and $\mu$ are kept constant. $N_{host}$ represents the amount of framework atoms and $\mu$ is the chemical potential of the solvent.\cite{Coudert2011} In this way, we take into account the possibility of framework expansion during the adsorption process.

These simulations require the chemical potential $\mu$ of the virtual reservoir in equilibrium with the system as an input.
Since the crystal and amorphous MOF systems are in equilibrium with liquid DMF at 400 K and 1 bar, the first step is to compute $\mu^0$ of DMF at these thermodynamic conditions.
To this end, we performed simulations of pure DMF at 400 K and different $\mu$ values and measured the equilibrium pressure $P$. The value of $\mu$ that gives $P=$ 1 bar is the DMF chemical potential in these conditions. A value of $\mu^0=$-104.5 kJ/mol was obtained.
The next step to obtain the adsorption free energy is to reversibly transform the empty framework into a filled framework at $\mu=\mu^0$. This is done by simulating the framework at different $\mu$ values and integrating the osmotic grand potential:

\begin{equation}
     \Delta\Omega^{OS} = - \int_{-\infty}^{\mu^0}\left\langle{N}\right\rangle_{(N_{host},T,P)} d\mu
\end{equation}
where $\left\langle{N}\right\rangle=0$ is the average amount of solvent molecules adsorbed. In practice, the lower boundary of the integral is replaced by a value of $\mu$ low enough to prevent guest adsorption.
The amorphous-to-crystal simulations that were carried out without solvent can be considered as part of the same thermodynamics ensemble, with $\mu=-\infty$, thus the results from both kinds of simulations can be compared.

\begin{figure}[H]
\centering
\includegraphics[width=0.55\textwidth]{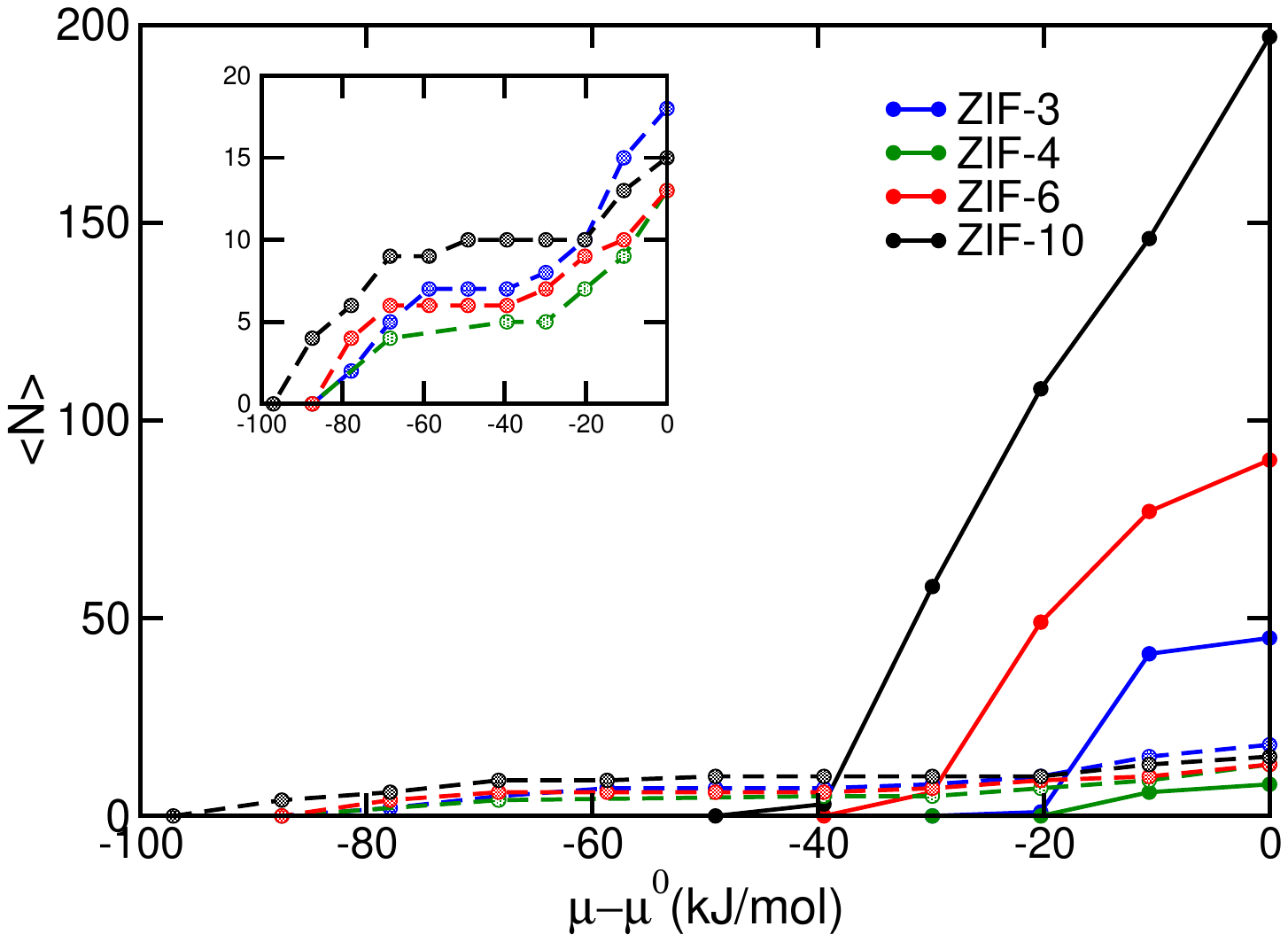}
\caption{\label{fig:gcmc}{DMF Adsorption curves for crystalline and amorphous ZIF phases. The inset shows only the amorphous phases for a clearer comparison.}}
\end{figure}

The $\left\langle{N}\right\rangle$ vs $\mu$ curves for all the polymorphs in their crystalline and amorphous states are shown in Figure \ref{fig:gcmc}.

\addcontentsline{toc}{section}{Classification algorithms}
\section{Classification algorithms}

\addcontentsline{toc}{subsection}{Chemical environment descriptors for neural network classifiers}
\subsection{Chemical environment descriptors for neural network classifiers}

We described the chemical environment of a Zn$^{2+}$ ion through Behler-Parrinello Symmetry Functions (BPSF).\cite{Behler2011}
These functions fulfil the basic criteria for well-behaved chemical descriptors, that is, they yield the same value for two configurations that are related in that one of them is the result of a translation, rotation or same-element-atom-permutation operation applied over the other. These atom-centred many-body functions can be classified into two types: \emph{radial} and \emph{angular}. The former ones are given by the sum over two-body terms and are related to the connectivity of the central atom, while in the latter ones, three-body terms are considered. We used 24 symmetry functions of the type:

\begin{equation}
G_i^{rad}=\sum_{j \neq i} e^{-\eta R_{ij}^2} \cdot f_c(R_{ij}) 
\label{eq:sfrad}
\end{equation}
\begin{equation}
G_i^{ang}=\sum_{j,k\neq i} (1-\lambda \cos{\theta_{ijk}}) \cdot e^{-\eta(R_{ij}^2+R_{ik}^2+R_{jk}^2)} \nonumber
\end{equation}
\begin{equation}
\cdot f_c(R_{ij}) \cdot f_c(R_{ik}) \cdot f_c(R_{jk})
\label{eq:sfang}
\end{equation}

\noindent
where $R_{ij}$ is the distance between atoms $i$ and $j$, $\theta_{ijk}$ is the angle between atoms $j$, $i$ and $k$ in that order, and $f_c$ is a cutoff function that decays to zero at a distance $R_c$. $\eta$ and $\lambda$ are parameters that change between different symmetry functions.
Each of the $n_{Zn^{2+}}$ cations in a given configuration will be thus characterised through 8 radial and 16 angular BPSF, half of them corresponding to Zn--Zn descriptors and the other half to Zn--N descriptors, which considers metal and ligand environments respectively. All other atom pairs were ignored.

The values of $\eta$ for the radial functions are 0.005, 0.0075, 0.01 and 0.02 Bohr$^{-2}$. With each of these $\eta$ values, we construct two angular functions with $\lambda$=$\pm 1$. The cutoff radius $R_c$ was set to 1.3 nm. The symmetry functions were normalised so that the highest and lowest values in the database were assigned one and zero respectively. 
The cutoff function $f_c$ was defined as:

\begin{equation}
    f_c(R) = \left\{ \begin{array}{lcc} \frac{1}{2}(cos(\frac{\pi R}{R_c})+1) & if & R \leq R_c \\ \\ 0 & if & R > R_c \end{array} \right.
\end{equation}

\addcontentsline{toc}{subsection}{Neural network training}
\subsection{Neural network training}

The training of the neural network was done using the Pytorch package implemented in Python.\cite{10.5555/3454287.3455008}
Rectified linear activation functions were employed for the hidden layer while the softmax function was used for the output layer.
For both PNC and amorphous classification, the neural network architecture comprised an input layer of 24 nodes, two hidden layers of 64 nodes, and an output layer of four nodes that represent the probabilities associated to each polymorph class.
The categorical cross-entropy was selected as loss function to be minimised.\cite{pml1Book}
We employed the Adam algorithm\cite{https://doi.org/10.48550/arxiv.1412.6980} for the optimisation of the neural network parameters. This process was done over 1000 epochs of the training set.

\addcontentsline{toc}{section}{Radial distribution functions of amorphous phases}
\section{Radial distribution functions of amorphous phases}

\begin{figure}[H]
\centering
\includegraphics[width=0.55\textwidth]{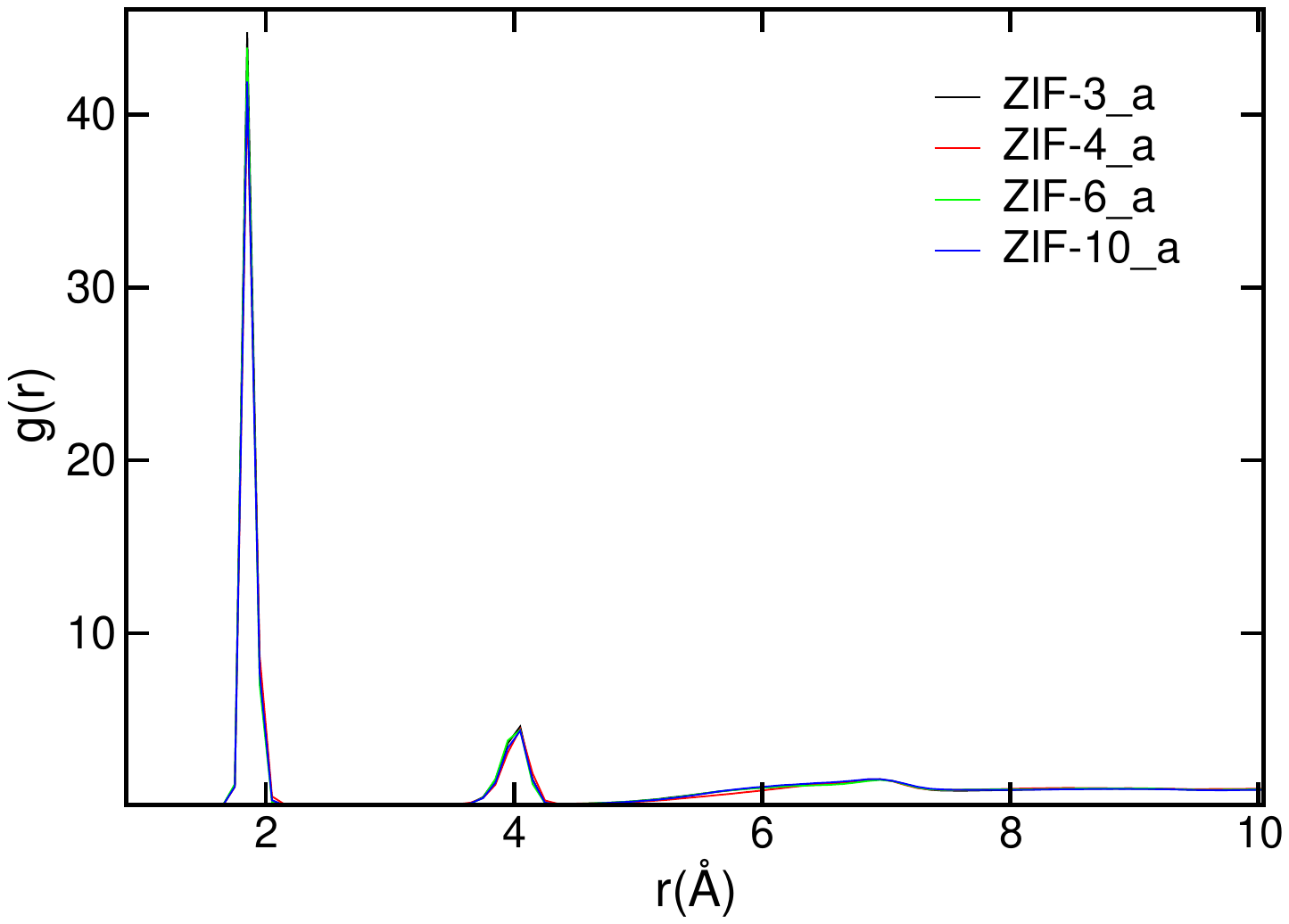}
\caption{\label{fig:gr}{Radial distribution functions of Zn--N pairs for the amorphous phases for each of the polymorphs studied.}}
\end{figure}

\addcontentsline{toc}{section}{Selected features of amorphous-to-crystal transitions}
\section{Selected features of amorphous-to-crystal transitions}

\begin{figure}[H]
\centering
\includegraphics[width=0.55\textwidth]{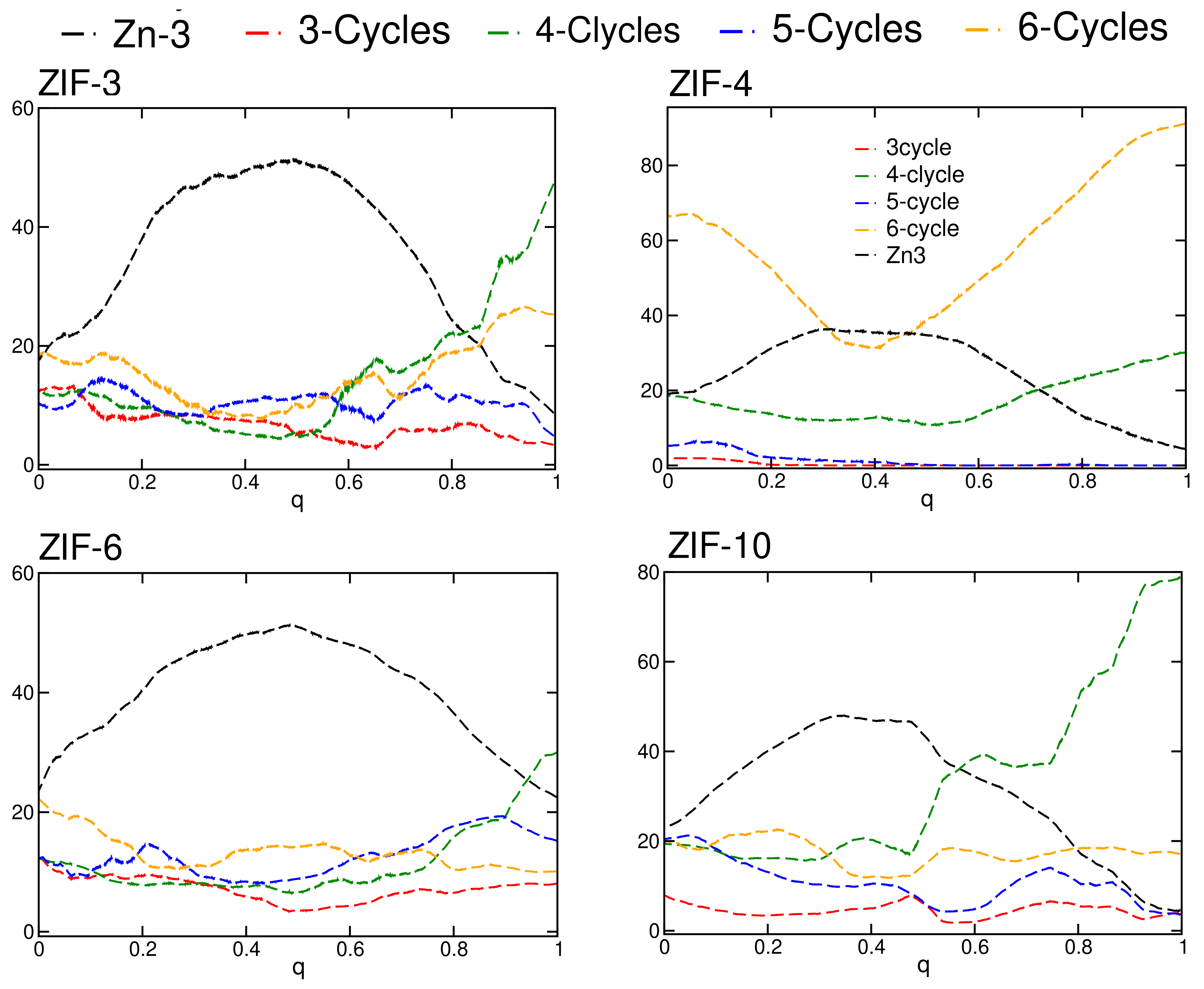}
\caption{\label{fig:psi_q}{Average values of selected geometrical features as a function of the reaction coordinate $q$ for each polymorph. The black curve corresponds to the number of 3-coordinated Zn$^{2+}$ ions. Red, green, blue and orange curves correspond to the number of 3-, 4-, 5- and 6-membered rings respectively.}}
\end{figure}

\addcontentsline{toc}{section}{$\Delta$G(q) of pore formation}
\section{$\Delta$G(q) of pore formation}

\begin{figure}[H]
\centering
\includegraphics[width=0.55\textwidth]{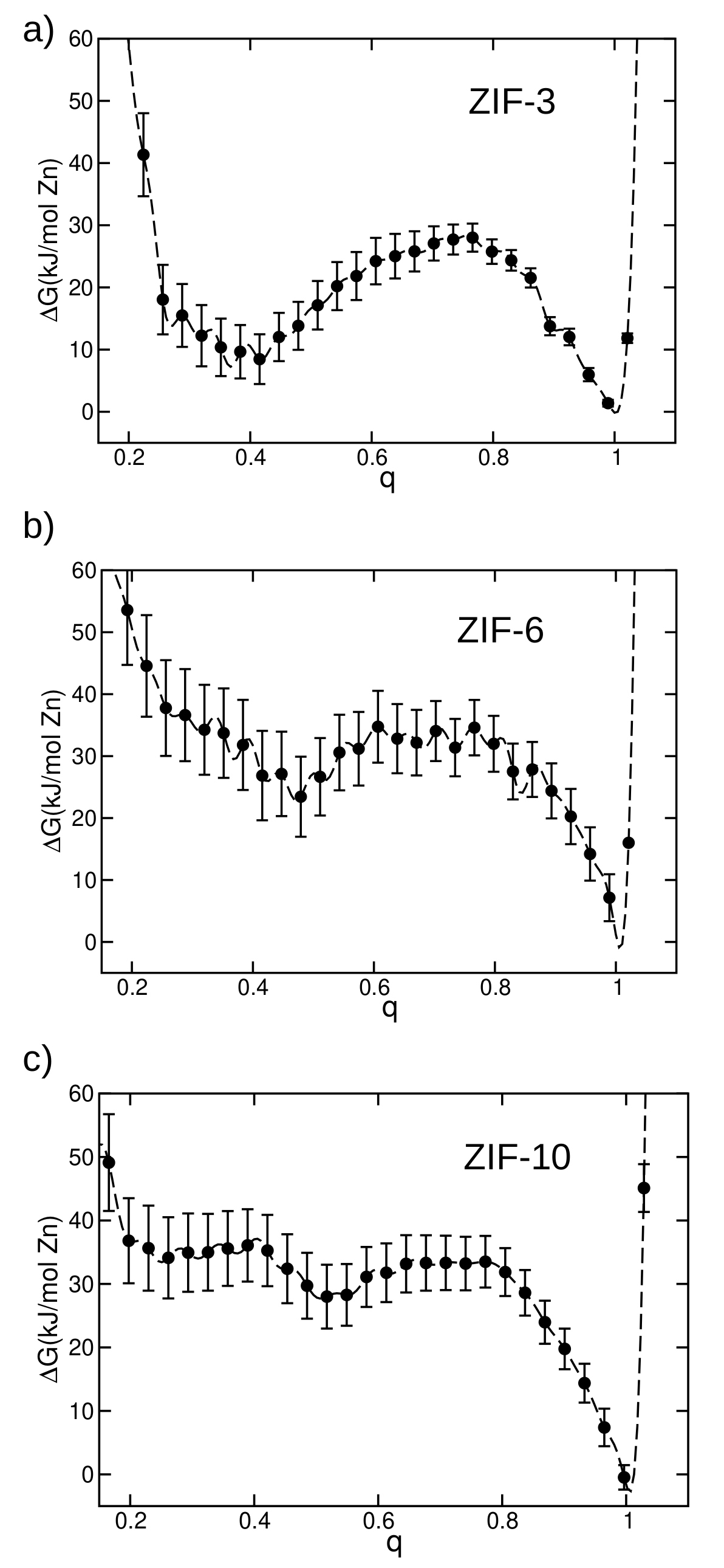}
\caption{\label{fig:gq}{Free energy per Zn$^{2+}$ ion as a function of the reaction coordinate $q$ for the formation of ZIF-3, ZIF-6 and ZIF-10 pores.}}
\end{figure}

\addcontentsline{toc}{section}{Pre-nucleation clusters snapshots}
\section{Pre-nucleation clusters snapshots}

\begin{figure}[H]
\centering
\includegraphics[width=0.55\textwidth]{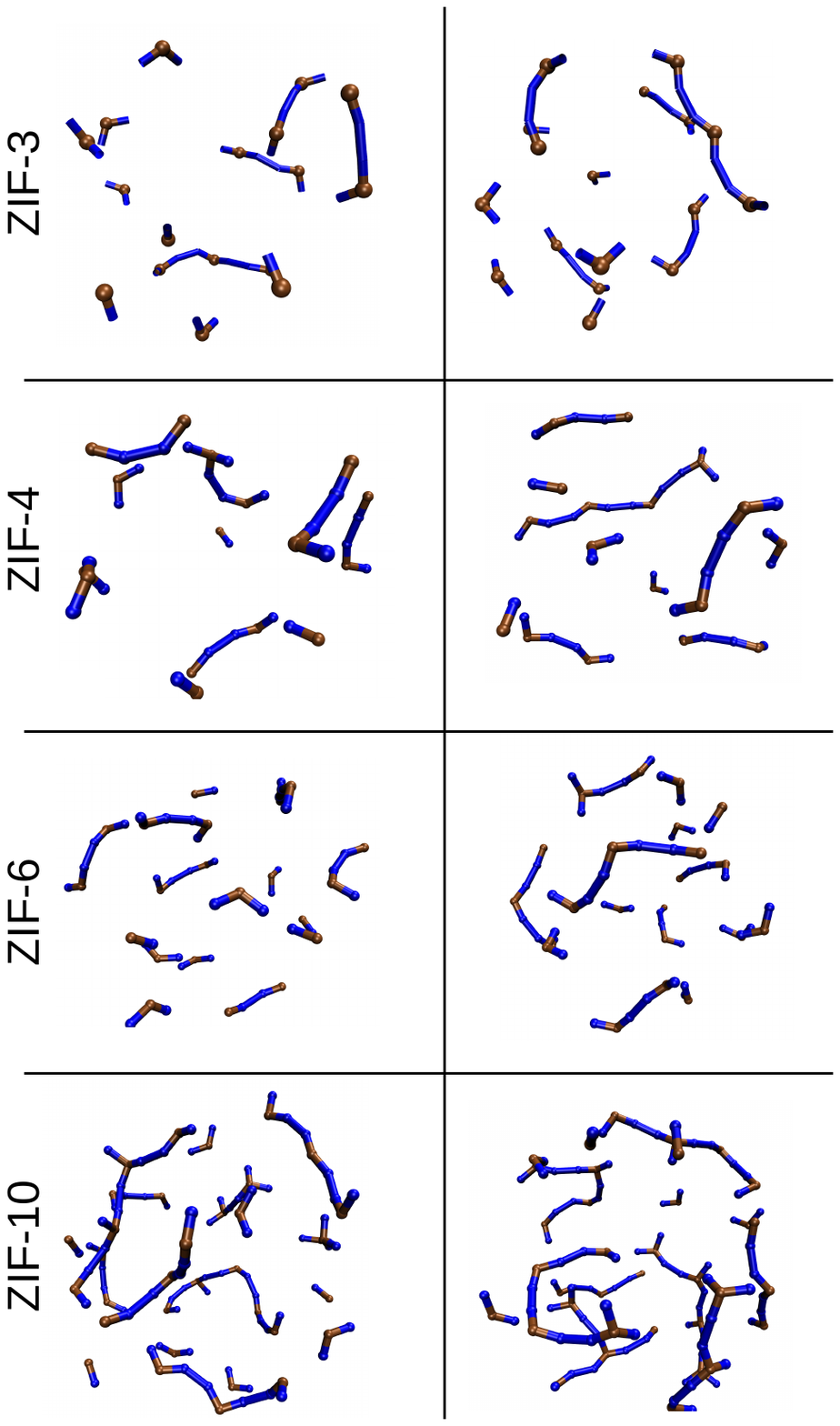}
\caption{\label{fig:pncs}{Additional snapshots for the pre-nucleation clusters associated to the formation of each polymorph pore.}}
\end{figure}

\addcontentsline{toc}{section}{Experimental synthesis conditions for ZIFs}
\section{Experimental synthesis conditions for ZIFs}

\begin{table}[H]
\small
  \caption{Experimental synthesis conditions for ZIF polymorphs taken from Ref. \citenum{Park2006}. Amount of metal and ligand, metal to ligand ratio (M/L), final volume $V$, temperature $T$, presence of additives and time $t$ are summarized.
  $^*$ ZIF-6 was obtained by combinatorial experiments}
  \label{tabla}
  \begin{tabular*}{0.9\textwidth}{@{\extracolsep{\fill}}llllllll}
    \hline
    & Zn$^{2+}$ (mol) & Im$^{-}$ (mol) & M/L ratio & $V$ (ml) & $T$ (K) & Additives & $t$(days)\\
    \hline
ZIF-3       & 3.82x10$^{-5}$    & 4.41x10$^{-4}$  & 1:11.5 & 4 & 358 & NPM (1 ml) & 4\\
ZIF-4    & 1.53x10$^{-4}$  & 4.41x10$^{-4}$  & 1:2.9 & 4 & 403 & - & 2\\
ZIF-6        & 3.23x10$^{-6}$  & 3.86x10$^{-5}$  & 1:12 & -$^*$ & 358 & -  & 3 \\
ZIF-10        & 3.82x10$^{-5}$  & 4.41x10$^{-4}$  & 1:11.5 & 4 & 358 & - & 4\\
    \hline
  \end{tabular*}
\end{table}

\bibliography{magnify}

\providecommand{\latin}[1]{#1}
\makeatletter
\providecommand{\doi}
  {\begingroup\let\do\@makeother\dospecials
  \catcode`\{=1 \catcode`\}=2 \doi@aux}
\providecommand{\doi@aux}[1]{\endgroup\texttt{#1}}
\makeatother
\providecommand*\mcitethebibliography{\thebibliography}
\csname @ifundefined\endcsname{endmcitethebibliography}  {\let\endmcitethebibliography\endthebibliography}{}
\begin{mcitethebibliography}{67}
\providecommand*\natexlab[1]{#1}
\providecommand*\mciteSetBstSublistMode[1]{}
\providecommand*\mciteSetBstMaxWidthForm[2]{}
\providecommand*\mciteBstWouldAddEndPuncttrue
  {\def\EndOfBibitem{\unskip.}}
\providecommand*\mciteBstWouldAddEndPunctfalse
  {\let\EndOfBibitem\relax}
\providecommand*\mciteSetBstMidEndSepPunct[3]{}
\providecommand*\mciteSetBstSublistLabelBeginEnd[3]{}
\providecommand*\EndOfBibitem{}
\mciteSetBstSublistMode{f}
\mciteSetBstMaxWidthForm{subitem}{(\alph{mcitesubitemcount})}
\mciteSetBstSublistLabelBeginEnd
  {\mcitemaxwidthsubitemform\space}
  {\relax}
  {\relax}

\bibitem[Meek \latin{et~al.}(2010)Meek, Greathouse, and Allendorf]{Meek2010}
Meek,~S.~T.; Greathouse,~J.~A.; Allendorf,~M.~D. Metal‐Organic Frameworks: A Rapidly Growing Class of Versatile Nanoporous Materials. \emph{Advanced Materials} \textbf{2010}, \emph{23}, 249–267\relax
\mciteBstWouldAddEndPuncttrue
\mciteSetBstMidEndSepPunct{\mcitedefaultmidpunct}
{\mcitedefaultendpunct}{\mcitedefaultseppunct}\relax
\EndOfBibitem
\bibitem[He \latin{et~al.}(2023)He, Lv, Guan, and Yu]{He2023}
He,~W.; Lv,~D.; Guan,~Y.; Yu,~S. Post-synthesis modification of metal–organic frameworks: synthesis, characteristics, and applications. \emph{Journal of Materials Chemistry A} \textbf{2023}, \emph{11}, 24519–24550\relax
\mciteBstWouldAddEndPuncttrue
\mciteSetBstMidEndSepPunct{\mcitedefaultmidpunct}
{\mcitedefaultendpunct}{\mcitedefaultseppunct}\relax
\EndOfBibitem
\bibitem[Kim \latin{et~al.}(2018)Kim, Rao, Kapustin, Zhao, Yang, Yaghi, and Wang]{Kim2018}
Kim,~H.; Rao,~S.~R.; Kapustin,~E.~A.; Zhao,~L.; Yang,~S.; Yaghi,~O.~M.; Wang,~E.~N. Adsorption-based atmospheric water harvesting device for arid climates. \emph{Nature Communications} \textbf{2018}, \emph{9}\relax
\mciteBstWouldAddEndPuncttrue
\mciteSetBstMidEndSepPunct{\mcitedefaultmidpunct}
{\mcitedefaultendpunct}{\mcitedefaultseppunct}\relax
\EndOfBibitem
\bibitem[Osterrieth and Fairen‐Jimenez(2021)Osterrieth, and Fairen‐Jimenez]{Osterrieth2020}
Osterrieth,~J. W.~M.; Fairen‐Jimenez,~D. Metal–Organic Framework Composites for Theragnostics and Drug Delivery Applications. \emph{Biotechnology Journal} \textbf{2021}, \emph{16}, 2000005\relax
\mciteBstWouldAddEndPuncttrue
\mciteSetBstMidEndSepPunct{\mcitedefaultmidpunct}
{\mcitedefaultendpunct}{\mcitedefaultseppunct}\relax
\EndOfBibitem
\bibitem[Mahajan and Lahtinen(2022)Mahajan, and Lahtinen]{Mahajan2022}
Mahajan,~S.; Lahtinen,~M. Recent progress in metal-organic frameworks (MOFs) for CO2 capture at different pressures. \emph{Journal of Environmental Chemical Engineering} \textbf{2022}, \emph{10}, 108930\relax
\mciteBstWouldAddEndPuncttrue
\mciteSetBstMidEndSepPunct{\mcitedefaultmidpunct}
{\mcitedefaultendpunct}{\mcitedefaultseppunct}\relax
\EndOfBibitem
\bibitem[Yaghi \latin{et~al.}(2003)Yaghi, O’Keeffe, Ockwig, Chae, Eddaoudi, and Kim]{Yaghi2003}
Yaghi,~O.~M.; O’Keeffe,~M.; Ockwig,~N.~W.; Chae,~H.~K.; Eddaoudi,~M.; Kim,~J. Reticular synthesis and the design of new materials. \emph{Nature} \textbf{2003}, \emph{423}, 705–714\relax
\mciteBstWouldAddEndPuncttrue
\mciteSetBstMidEndSepPunct{\mcitedefaultmidpunct}
{\mcitedefaultendpunct}{\mcitedefaultseppunct}\relax
\EndOfBibitem
\bibitem[Chen \latin{et~al.}(2022)Chen, Kirlikovali, Li, and Farha]{Chen2022}
Chen,~Z.; Kirlikovali,~K.~O.; Li,~P.; Farha,~O.~K. Reticular Chemistry for Highly Porous Metal–Organic Frameworks: The Chemistry and Applications. \emph{Accounts of Chemical Research} \textbf{2022}, \emph{55}, 579–591\relax
\mciteBstWouldAddEndPuncttrue
\mciteSetBstMidEndSepPunct{\mcitedefaultmidpunct}
{\mcitedefaultendpunct}{\mcitedefaultseppunct}\relax
\EndOfBibitem
\bibitem[Guillerm and Eddaoudi(2021)Guillerm, and Eddaoudi]{Guillerm2021}
Guillerm,~V.; Eddaoudi,~M. The Importance of Highly Connected Building Units in Reticular Chemistry: Thoughtful Design of Metal–Organic Frameworks. \emph{Accounts of Chemical Research} \textbf{2021}, \emph{54}, 3298–3312\relax
\mciteBstWouldAddEndPuncttrue
\mciteSetBstMidEndSepPunct{\mcitedefaultmidpunct}
{\mcitedefaultendpunct}{\mcitedefaultseppunct}\relax
\EndOfBibitem
\bibitem[Freund \latin{et~al.}(2021)Freund, Canossa, Cohen, Yan, Deng, Guillerm, Eddaoudi, Madden, Fairen‐Jimenez, Lyu, Macreadie, Ji, Zhang, Wang, Haase, W\"{o}ll, Zaremba, Andreo, Wuttke, and Diercks]{Freund2021}
Freund,~R. \latin{et~al.}  25 Years of Reticular Chemistry. \emph{Angewandte Chemie International Edition} \textbf{2021}, \emph{60}, 23946–23974\relax
\mciteBstWouldAddEndPuncttrue
\mciteSetBstMidEndSepPunct{\mcitedefaultmidpunct}
{\mcitedefaultendpunct}{\mcitedefaultseppunct}\relax
\EndOfBibitem
\bibitem[Kalmutzki \latin{et~al.}(2018)Kalmutzki, Hanikel, and Yaghi]{Kalmutzki2018}
Kalmutzki,~M.~J.; Hanikel,~N.; Yaghi,~O.~M. Secondary building units as the turning point in the development of the reticular chemistry of MOFs. \emph{Science Advances} \textbf{2018}, \emph{4}, eaat9180\relax
\mciteBstWouldAddEndPuncttrue
\mciteSetBstMidEndSepPunct{\mcitedefaultmidpunct}
{\mcitedefaultendpunct}{\mcitedefaultseppunct}\relax
\EndOfBibitem
\bibitem[Jiang \latin{et~al.}(2021)Jiang, Alezi, and Eddaoudi]{Jiang2021}
Jiang,~H.; Alezi,~D.; Eddaoudi,~M. A reticular chemistry guide for the design of periodic solids. \emph{Nature Reviews Materials} \textbf{2021}, \emph{6}, 466–487\relax
\mciteBstWouldAddEndPuncttrue
\mciteSetBstMidEndSepPunct{\mcitedefaultmidpunct}
{\mcitedefaultendpunct}{\mcitedefaultseppunct}\relax
\EndOfBibitem
\bibitem[Stock and Biswas(2011)Stock, and Biswas]{Stock2011}
Stock,~N.; Biswas,~S. Synthesis of Metal-Organic Frameworks (MOFs): Routes to Various MOF Topologies, Morphologies, and Composites. \emph{Chemical Reviews} \textbf{2011}, \emph{112}, 933–969\relax
\mciteBstWouldAddEndPuncttrue
\mciteSetBstMidEndSepPunct{\mcitedefaultmidpunct}
{\mcitedefaultendpunct}{\mcitedefaultseppunct}\relax
\EndOfBibitem
\bibitem[Guillerm and Maspoch(2019)Guillerm, and Maspoch]{Guillerm2019}
Guillerm,~V.; Maspoch,~D. Geometry Mismatch and Reticular Chemistry: Strategies To Assemble Metal–Organic Frameworks with Non-default Topologies. \emph{Journal of the American Chemical Society} \textbf{2019}, \emph{141}, 16517–16538\relax
\mciteBstWouldAddEndPuncttrue
\mciteSetBstMidEndSepPunct{\mcitedefaultmidpunct}
{\mcitedefaultendpunct}{\mcitedefaultseppunct}\relax
\EndOfBibitem
\bibitem[Kitagawa \latin{et~al.}(2022)Kitagawa, Kaskel, and Xu]{Kitagawa2022}
Kitagawa,~S.; Kaskel,~S.; Xu,~Q. Metal‐Organic Frameworks: Synthesis, Structures, and Applications. \emph{Small Structures} \textbf{2022}, \emph{3}\relax
\mciteBstWouldAddEndPuncttrue
\mciteSetBstMidEndSepPunct{\mcitedefaultmidpunct}
{\mcitedefaultendpunct}{\mcitedefaultseppunct}\relax
\EndOfBibitem
\bibitem[Moosavi \latin{et~al.}(2019)Moosavi, Chidambaram, Talirz, Haranczyk, Stylianou, and Smit]{Moosavi2019}
Moosavi,~S.~M.; Chidambaram,~A.; Talirz,~L.; Haranczyk,~M.; Stylianou,~K.~C.; Smit,~B. Capturing chemical intuition in synthesis of metal-organic frameworks. \emph{Nature Communications} \textbf{2019}, \emph{10}\relax
\mciteBstWouldAddEndPuncttrue
\mciteSetBstMidEndSepPunct{\mcitedefaultmidpunct}
{\mcitedefaultendpunct}{\mcitedefaultseppunct}\relax
\EndOfBibitem
\bibitem[Zheng \latin{et~al.}(2025)Zheng, Rampal, Inizan, Borgs, Chayes, and Yaghi]{Zheng2025}
Zheng,~Z.; Rampal,~N.; Inizan,~T.~J.; Borgs,~C.; Chayes,~J.~T.; Yaghi,~O.~M. Large language models for reticular chemistry. \emph{Nature Reviews Materials} \textbf{2025}, \emph{10}, 369–381\relax
\mciteBstWouldAddEndPuncttrue
\mciteSetBstMidEndSepPunct{\mcitedefaultmidpunct}
{\mcitedefaultendpunct}{\mcitedefaultseppunct}\relax
\EndOfBibitem
\bibitem[Zhao \latin{et~al.}(2025)Zhao, Zhao, Wang, and Wang]{Zhao2025}
Zhao,~Y.; Zhao,~Y.; Wang,~J.; Wang,~Z. Artificial Intelligence Meets Laboratory Automation in Discovery and Synthesis of Metal–Organic Frameworks: A Review. \emph{Industrial \& Engineering Chemistry Research} \textbf{2025}, \emph{64}, 4637–4668\relax
\mciteBstWouldAddEndPuncttrue
\mciteSetBstMidEndSepPunct{\mcitedefaultmidpunct}
{\mcitedefaultendpunct}{\mcitedefaultseppunct}\relax
\EndOfBibitem
\bibitem[Zheng \latin{et~al.}(2023)Zheng, Zhang, Nguyen, Rampal, Alawadhi, Rong, Head-Gordon, Borgs, Chayes, and Yaghi]{Zheng2023}
Zheng,~Z.; Zhang,~O.; Nguyen,~H.~L.; Rampal,~N.; Alawadhi,~A.~H.; Rong,~Z.; Head-Gordon,~T.; Borgs,~C.; Chayes,~J.~T.; Yaghi,~O.~M. ChatGPT Research Group for Optimizing the Crystallinity of MOFs and COFs. \emph{ACS Central Science} \textbf{2023}, \emph{9}, 2161–2170\relax
\mciteBstWouldAddEndPuncttrue
\mciteSetBstMidEndSepPunct{\mcitedefaultmidpunct}
{\mcitedefaultendpunct}{\mcitedefaultseppunct}\relax
\EndOfBibitem
\bibitem[Park \latin{et~al.}(2006)Park, Ni, C{\^{o}}t{\'{e}}, Choi, Huang, Uribe-Romo, Chae, O'Keeffe, and Yaghi]{Park2006}
Park,~K.~S.; Ni,~Z.; C{\^{o}}t{\'{e}},~A.~P.; Choi,~J.~Y.; Huang,~R.; Uribe-Romo,~F.~J.; Chae,~H.~K.; O'Keeffe,~M.; Yaghi,~O.~M. Exceptional chemical and thermal stability of zeolitic imidazolate frameworks. \emph{Proceedings of the National Academy of Sciences} \textbf{2006}, \emph{103}, 10186--10191\relax
\mciteBstWouldAddEndPuncttrue
\mciteSetBstMidEndSepPunct{\mcitedefaultmidpunct}
{\mcitedefaultendpunct}{\mcitedefaultseppunct}\relax
\EndOfBibitem
\bibitem[Venna \latin{et~al.}(2010)Venna, Jasinski, and Carreon]{Venna2010}
Venna,~S.~R.; Jasinski,~J.~B.; Carreon,~M.~A. Structural Evolution of Zeolitic Imidazolate Framework-8. \emph{Journal of the American Chemical Society} \textbf{2010}, \emph{132}, 18030–18033\relax
\mciteBstWouldAddEndPuncttrue
\mciteSetBstMidEndSepPunct{\mcitedefaultmidpunct}
{\mcitedefaultendpunct}{\mcitedefaultseppunct}\relax
\EndOfBibitem
\bibitem[Bustamante \latin{et~al.}(2014)Bustamante, Fernández, and Zamaro]{Bustamante2014}
Bustamante,~E.~L.; Fernández,~J.~L.; Zamaro,~J.~M. Influence of the solvent in the synthesis of zeolitic imidazolate framework-8 (ZIF-8) nanocrystals at room temperature. \emph{Journal of Colloid and Interface Science} \textbf{2014}, \emph{424}, 37–43\relax
\mciteBstWouldAddEndPuncttrue
\mciteSetBstMidEndSepPunct{\mcitedefaultmidpunct}
{\mcitedefaultendpunct}{\mcitedefaultseppunct}\relax
\EndOfBibitem
\bibitem[Balog \latin{et~al.}(2022)Balog, Varga, Kukovecz, Tóth, Horváth, Lagzi, and Schuszter]{Balog2022}
Balog,~E.; Varga,~G.; Kukovecz,~A.; Tóth,~A.; Horváth,~D.; Lagzi,~I.; Schuszter,~G. Polymorph Selection of Zeolitic Imidazolate Frameworks via Kinetic and Thermodynamic Control. \emph{Crystal Growth \& Design} \textbf{2022}, \emph{22}, 4268–4276\relax
\mciteBstWouldAddEndPuncttrue
\mciteSetBstMidEndSepPunct{\mcitedefaultmidpunct}
{\mcitedefaultendpunct}{\mcitedefaultseppunct}\relax
\EndOfBibitem
\bibitem[Moh \latin{et~al.}(2013)Moh, Brenda, Anderson, and Attfield]{Moh2013}
Moh,~P.~Y.; Brenda,~M.; Anderson,~M.~W.; Attfield,~M.~P. Crystallisation of solvothermally synthesised ZIF-8 investigated at the bulk, single crystal and surface level. \emph{CrystEngComm} \textbf{2013}, \emph{15}, 9672\relax
\mciteBstWouldAddEndPuncttrue
\mciteSetBstMidEndSepPunct{\mcitedefaultmidpunct}
{\mcitedefaultendpunct}{\mcitedefaultseppunct}\relax
\EndOfBibitem
\bibitem[Cravillon \latin{et~al.}(2012)Cravillon, Schr\"{o}der, Bux, Rothkirch, Caro, and Wiebcke]{Cravillon2012}
Cravillon,~J.; Schr\"{o}der,~C.~A.; Bux,~H.; Rothkirch,~A.; Caro,~J.; Wiebcke,~M. Formate modulated solvothermal synthesis of ZIF-8 investigated using time-resolved in situ X-ray diffraction and scanning electron microscopy. \emph{CrystEngComm} \textbf{2012}, \emph{14}, 492–498\relax
\mciteBstWouldAddEndPuncttrue
\mciteSetBstMidEndSepPunct{\mcitedefaultmidpunct}
{\mcitedefaultendpunct}{\mcitedefaultseppunct}\relax
\EndOfBibitem
\bibitem[Jin \latin{et~al.}(2023)Jin, Wang, Boglaienko, Zhang, Zhao, Ma, Zhang, and De~Yoreo]{Jin2023}
Jin,~B.; Wang,~S.; Boglaienko,~D.; Zhang,~Z.; Zhao,~Q.; Ma,~X.; Zhang,~X.; De~Yoreo,~J.~J. The role of amorphous ZIF in ZIF-8 crystallization kinetics and morphology. \emph{Journal of Crystal Growth} \textbf{2023}, \emph{603}, 126989\relax
\mciteBstWouldAddEndPuncttrue
\mciteSetBstMidEndSepPunct{\mcitedefaultmidpunct}
{\mcitedefaultendpunct}{\mcitedefaultseppunct}\relax
\EndOfBibitem
\bibitem[Talosig \latin{et~al.}(2024)Talosig, Wang, Mulvey, Carpenter, Olivas, Katz, Zhu, and Patterson]{Talosig2024}
Talosig,~A.~R.; Wang,~F.; Mulvey,~J.~T.; Carpenter,~B.~P.; Olivas,~E.~M.; Katz,~B.~B.; Zhu,~C.; Patterson,~J.~P. Understanding the Nucleation and Growth of ZIF-8 Polymorphs. \emph{Crystal Growth \& Design} \textbf{2024}, \emph{24}, 4136–4142\relax
\mciteBstWouldAddEndPuncttrue
\mciteSetBstMidEndSepPunct{\mcitedefaultmidpunct}
{\mcitedefaultendpunct}{\mcitedefaultseppunct}\relax
\EndOfBibitem
\bibitem[Dok \latin{et~al.}(2025)Dok, Radhakrishnan, de~Jong, Becquevort, Deschaume, Chandran, de~Coene, Bartic, Van~der Auweraer, Thielemans, Kirschhock, van~der Veen, Verbiest, Breynaert, and Van~Cleuvenbergen]{Dok2025}
Dok,~A.~R.; Radhakrishnan,~S.; de~Jong,~F.; Becquevort,~E.; Deschaume,~O.; Chandran,~C.~V.; de~Coene,~Y.; Bartic,~C.; Van~der Auweraer,~M.; Thielemans,~W.; Kirschhock,~C.; van~der Veen,~M.~A.; Verbiest,~T.; Breynaert,~E.; Van~Cleuvenbergen,~S. Amorphous-to-Crystalline Transformation: How Cluster Aggregation Drives the Multistep Nucleation of ZIF-8. \emph{Journal of the American Chemical Society} \textbf{2025}, \emph{147}, 8455–8466\relax
\mciteBstWouldAddEndPuncttrue
\mciteSetBstMidEndSepPunct{\mcitedefaultmidpunct}
{\mcitedefaultendpunct}{\mcitedefaultseppunct}\relax
\EndOfBibitem
\bibitem[Filez \latin{et~al.}(2021)Filez, Caratelli, Rivera-Torrente, Muniz-Miranda, Hoek, Altelaar, Heck, Van~Speybroeck, and Weckhuysen]{Filez2021}
Filez,~M.; Caratelli,~C.; Rivera-Torrente,~M.; Muniz-Miranda,~F.; Hoek,~M.; Altelaar,~M.; Heck,~A.~J.; Van~Speybroeck,~V.; Weckhuysen,~B.~M. Elucidation of the pre-nucleation phase directing metal-organic framework formation. \emph{Cell Reports Physical Science} \textbf{2021}, \emph{2}, 100680\relax
\mciteBstWouldAddEndPuncttrue
\mciteSetBstMidEndSepPunct{\mcitedefaultmidpunct}
{\mcitedefaultendpunct}{\mcitedefaultseppunct}\relax
\EndOfBibitem
\bibitem[Balestra and Semino(2022)Balestra, and Semino]{Balestra2022}
Balestra,~S. R.~G.; Semino,~R. Computer simulation of the early stages of self-assembly and thermal decomposition of {ZIF}-8. \emph{The Journal of Chemical Physics} \textbf{2022}, \emph{157}, 184502\relax
\mciteBstWouldAddEndPuncttrue
\mciteSetBstMidEndSepPunct{\mcitedefaultmidpunct}
{\mcitedefaultendpunct}{\mcitedefaultseppunct}\relax
\EndOfBibitem
\bibitem[Balestra \latin{et~al.}(2023)Balestra, Martínez-Haya, Cruz-Hernández, Lewis, Woodley, Semino, Maurin, Ruiz-Salvador, and Hamad]{Balestra2023}
Balestra,~S. R.~G.; Martínez-Haya,~B.; Cruz-Hernández,~N.; Lewis,~D.~W.; Woodley,~S.~M.; Semino,~R.; Maurin,~G.; Ruiz-Salvador,~A.~R.; Hamad,~S. Nucleation of zeolitic imidazolate frameworks: from molecules to nanoparticles. \emph{Nanoscale} \textbf{2023}, \emph{15}, 3504–3519\relax
\mciteBstWouldAddEndPuncttrue
\mciteSetBstMidEndSepPunct{\mcitedefaultmidpunct}
{\mcitedefaultendpunct}{\mcitedefaultseppunct}\relax
\EndOfBibitem
\bibitem[Méndez and Semino(2025)Méndez, and Semino]{Mendez2025}
Méndez,~E.; Semino,~R. Thermodynamic insights into the self-assembly of zeolitic imidazolate frameworks from computer simulations. \emph{Chem. Sci.} \textbf{2025}, \emph{16}, 11979--11988\relax
\mciteBstWouldAddEndPuncttrue
\mciteSetBstMidEndSepPunct{\mcitedefaultmidpunct}
{\mcitedefaultendpunct}{\mcitedefaultseppunct}\relax
\EndOfBibitem
\bibitem[Gargari and Semino(2025)Gargari, and Semino]{AndarziGargari2025}
Gargari,~S.~A.; Semino,~R. Unveiling ZIF-8 Nucleation Mechanisms through Molecular Simulation: Role of Temperature, Solvent, and Reactant Concentration. \emph{Chemistry of Materials} \textbf{2025}, \emph{37}, 9460–9470\relax
\mciteBstWouldAddEndPuncttrue
\mciteSetBstMidEndSepPunct{\mcitedefaultmidpunct}
{\mcitedefaultendpunct}{\mcitedefaultseppunct}\relax
\EndOfBibitem
\bibitem[Widmer \latin{et~al.}(2019)Widmer, Lampronti, Chibani, Wilson, Anzellini, Farsang, Kleppe, Casati, MacLeod, Redfern, Coudert, and Bennett]{Widmer2019}
Widmer,~R.~N.; Lampronti,~G.~I.; Chibani,~S.; Wilson,~C.~W.; Anzellini,~S.; Farsang,~S.; Kleppe,~A.~K.; Casati,~N. P.~M.; MacLeod,~S.~G.; Redfern,~S. A.~T.; Coudert,~F.-X.; Bennett,~T.~D. Rich Polymorphism of a Metal{\textendash}Organic Framework in Pressure{\textendash}Temperature Space. \emph{Journal of the American Chemical Society} \textbf{2019}, \emph{141}, 9330--9337\relax
\mciteBstWouldAddEndPuncttrue
\mciteSetBstMidEndSepPunct{\mcitedefaultmidpunct}
{\mcitedefaultendpunct}{\mcitedefaultseppunct}\relax
\EndOfBibitem
\bibitem[Laio and Parrinello(2002)Laio, and Parrinello]{Laio2002}
Laio,~A.; Parrinello,~M. Escaping free-energy minima. \emph{Proceedings of the National Academy of Sciences} \textbf{2002}, \emph{99}, 12562–12566\relax
\mciteBstWouldAddEndPuncttrue
\mciteSetBstMidEndSepPunct{\mcitedefaultmidpunct}
{\mcitedefaultendpunct}{\mcitedefaultseppunct}\relax
\EndOfBibitem
\bibitem[Díaz~Leines and Ensing(2012)Díaz~Leines, and Ensing]{DazLeines2012}
Díaz~Leines,~G.; Ensing,~B. Path Finding on High-Dimensional Free Energy Landscapes. \emph{Physical Review Letters} \textbf{2012}, \emph{109}\relax
\mciteBstWouldAddEndPuncttrue
\mciteSetBstMidEndSepPunct{\mcitedefaultmidpunct}
{\mcitedefaultendpunct}{\mcitedefaultseppunct}\relax
\EndOfBibitem
\bibitem[Méndez and Semino(2024)Méndez, and Semino]{Mendez2024}
Méndez,~E.; Semino,~R. Microscopic mechanism of thermal amorphization of ZIF-4 and melting of ZIF-zni revealed via molecular dynamics and machine learning techniques. \emph{Journal of Materials Chemistry A} \textbf{2024}, \emph{12}, 4572–4582\relax
\mciteBstWouldAddEndPuncttrue
\mciteSetBstMidEndSepPunct{\mcitedefaultmidpunct}
{\mcitedefaultendpunct}{\mcitedefaultseppunct}\relax
\EndOfBibitem
\bibitem[Bouëssel~du Bourg \latin{et~al.}(2014)Bouëssel~du Bourg, Ortiz, Boutin, and Coudert]{BousselduBourg2014}
Bouëssel~du Bourg,~L.; Ortiz,~A.~U.; Boutin,~A.; Coudert,~F.-X. Thermal and mechanical stability of zeolitic imidazolate frameworks polymorphs. \emph{APL Materials} \textbf{2014}, \emph{2}, 124110\relax
\mciteBstWouldAddEndPuncttrue
\mciteSetBstMidEndSepPunct{\mcitedefaultmidpunct}
{\mcitedefaultendpunct}{\mcitedefaultseppunct}\relax
\EndOfBibitem
\bibitem[Coudert \latin{et~al.}(2011)Coudert, Boutin, Jeffroy, Mellot‐Draznieks, and Fuchs]{Coudert2011}
Coudert,~F.; Boutin,~A.; Jeffroy,~M.; Mellot‐Draznieks,~C.; Fuchs,~A.~H. Thermodynamic Methods and Models to Study Flexible Metal–Organic Frameworks. \emph{ChemPhysChem} \textbf{2011}, \emph{12}, 247–258\relax
\mciteBstWouldAddEndPuncttrue
\mciteSetBstMidEndSepPunct{\mcitedefaultmidpunct}
{\mcitedefaultendpunct}{\mcitedefaultseppunct}\relax
\EndOfBibitem
\bibitem[Rogge \latin{et~al.}(2019)Rogge, Goeminne, Demuynck, Gutiérrez‐Sevillano, Vandenbrande, Vanduyfhuys, Waroquier, Verstraelen, and Van~Speybroeck]{Rogge2019_2}
Rogge,~S. M.~J.; Goeminne,~R.; Demuynck,~R.; Gutiérrez‐Sevillano,~J.~J.; Vandenbrande,~S.; Vanduyfhuys,~L.; Waroquier,~M.; Verstraelen,~T.; Van~Speybroeck,~V. Modeling Gas Adsorption in Flexible Metal–Organic Frameworks via Hybrid Monte Carlo/Molecular Dynamics Schemes. \emph{Advanced Theory and Simulations} \textbf{2019}, \emph{2}\relax
\mciteBstWouldAddEndPuncttrue
\mciteSetBstMidEndSepPunct{\mcitedefaultmidpunct}
{\mcitedefaultendpunct}{\mcitedefaultseppunct}\relax
\EndOfBibitem
\bibitem[Jeffroy \latin{et~al.}(2008)Jeffroy, Fuchs, and Boutin]{Jeffroy2008}
Jeffroy,~M.; Fuchs,~A.~H.; Boutin,~A. Structural changes in nanoporous solids due to fluid adsorption: thermodynamic analysis and Monte Carlo simulations. \emph{Chemical Communications} \textbf{2008}, 3275\relax
\mciteBstWouldAddEndPuncttrue
\mciteSetBstMidEndSepPunct{\mcitedefaultmidpunct}
{\mcitedefaultendpunct}{\mcitedefaultseppunct}\relax
\EndOfBibitem
\bibitem[Behler(2011)]{Behler2011}
Behler,~J. Atom-centered symmetry functions for constructing high-dimensional neural network potentials. \emph{The Journal of Chemical Physics} \textbf{2011}, \emph{134}, 074106\relax
\mciteBstWouldAddEndPuncttrue
\mciteSetBstMidEndSepPunct{\mcitedefaultmidpunct}
{\mcitedefaultendpunct}{\mcitedefaultseppunct}\relax
\EndOfBibitem
\bibitem[Andarzi~Gargari \latin{et~al.}(2026)Andarzi~Gargari, Méndez, and Semino]{AndarziGargari2026}
Andarzi~Gargari,~S.; Méndez,~E.; Semino,~R. Computer Simulation of the Growth of a Metal–Organic Framework Proto-Crystal at Constant Chemical Potential. \emph{ACS Applied Nano Materials} \textbf{2026}, \relax
\mciteBstWouldAddEndPunctfalse
\mciteSetBstMidEndSepPunct{\mcitedefaultmidpunct}
{}{\mcitedefaultseppunct}\relax
\EndOfBibitem
\bibitem[Karmakar \latin{et~al.}(2019)Karmakar, Piaggi, and Parrinello]{Karmakar2019}
Karmakar,~T.; Piaggi,~P.~M.; Parrinello,~M. Molecular Dynamics Simulations of Crystal Nucleation from Solution at Constant Chemical Potential. \emph{Journal of Chemical Theory and Computation} \textbf{2019}, \emph{15}, 6923–6930\relax
\mciteBstWouldAddEndPuncttrue
\mciteSetBstMidEndSepPunct{\mcitedefaultmidpunct}
{\mcitedefaultendpunct}{\mcitedefaultseppunct}\relax
\EndOfBibitem
\bibitem[Neha \latin{et~al.}(2022)Neha, Tiwari, Mondal, Kumari, and Karmakar]{Neha2022}
Neha; Tiwari,~V.; Mondal,~S.; Kumari,~N.; Karmakar,~T. Collective Variables for Crystallization Simulations - from Early Developments to Recent Advances. \emph{ACS Omega} \textbf{2022}, \emph{8}, 127–146\relax
\mciteBstWouldAddEndPuncttrue
\mciteSetBstMidEndSepPunct{\mcitedefaultmidpunct}
{\mcitedefaultendpunct}{\mcitedefaultseppunct}\relax
\EndOfBibitem
\bibitem[Van~Vleet \latin{et~al.}(2018)Van~Vleet, Weng, Li, and Schmidt]{VanVleet2018}
Van~Vleet,~M.~J.; Weng,~T.; Li,~X.; Schmidt,~J. In Situ, Time-Resolved, and Mechanistic Studies of Metal–Organic Framework Nucleation and Growth. \emph{Chemical Reviews} \textbf{2018}, \emph{118}, 3681–3721\relax
\mciteBstWouldAddEndPuncttrue
\mciteSetBstMidEndSepPunct{\mcitedefaultmidpunct}
{\mcitedefaultendpunct}{\mcitedefaultseppunct}\relax
\EndOfBibitem
\bibitem[Lewis \latin{et~al.}(2009)Lewis, Ruiz-Salvador, Gómez, Rodriguez-Albelo, Coudert, Slater, Cheetham, and Mellot-Draznieks]{Lewis2009}
Lewis,~D.~W.; Ruiz-Salvador,~A.~R.; Gómez,~A.; Rodriguez-Albelo,~L.~M.; Coudert,~F.-X.; Slater,~B.; Cheetham,~A.~K.; Mellot-Draznieks,~C. Zeolitic imidazole frameworks: structural and energetics trends compared with their zeolite analogues. \emph{CrystEngComm} \textbf{2009}, \emph{11}, 2272\relax
\mciteBstWouldAddEndPuncttrue
\mciteSetBstMidEndSepPunct{\mcitedefaultmidpunct}
{\mcitedefaultendpunct}{\mcitedefaultseppunct}\relax
\EndOfBibitem
\bibitem[D\"{u}rholt \latin{et~al.}(2019)D\"{u}rholt, Fraux, Coudert, and Schmid]{Drholt2019}
D\"{u}rholt,~J.~P.; Fraux,~G.; Coudert,~F.-X.; Schmid,~R. Ab Initio Derived Force Fields for Zeolitic Imidazolate Frameworks: MOF-FF for ZIFs. \emph{Journal of Chemical Theory and Computation} \textbf{2019}, \emph{15}, 2420–2432\relax
\mciteBstWouldAddEndPuncttrue
\mciteSetBstMidEndSepPunct{\mcitedefaultmidpunct}
{\mcitedefaultendpunct}{\mcitedefaultseppunct}\relax
\EndOfBibitem
\bibitem[Thompson \latin{et~al.}(2022)Thompson, Aktulga, Berger, Bolintineanu, Brown, Crozier, in~{\textquotesingle}t~Veld, Kohlmeyer, Moore, Nguyen, Shan, Stevens, Tranchida, Trott, and Plimpton]{lammps}
Thompson,~A.~P.; Aktulga,~H.~M.; Berger,~R.; Bolintineanu,~D.~S.; Brown,~W.~M.; Crozier,~P.~S.; in~{\textquotesingle}t~Veld,~P.~J.; Kohlmeyer,~A.; Moore,~S.~G.; Nguyen,~T.~D.; Shan,~R.; Stevens,~M.~J.; Tranchida,~J.; Trott,~C.; Plimpton,~S.~J. {LAMMPS} - a flexible simulation tool for particle-based materials modeling at the atomic, meso, and continuum scales. \emph{Computer Physics Communications} \textbf{2022}, \emph{271}, 108171\relax
\mciteBstWouldAddEndPuncttrue
\mciteSetBstMidEndSepPunct{\mcitedefaultmidpunct}
{\mcitedefaultendpunct}{\mcitedefaultseppunct}\relax
\EndOfBibitem
\bibitem[Tribello \latin{et~al.}(2014)Tribello, Bonomi, Branduardi, Camilloni, and Bussi]{Tribello2014}
Tribello,~G.~A.; Bonomi,~M.; Branduardi,~D.; Camilloni,~C.; Bussi,~G. PLUMED 2: New feathers for an old bird. \emph{Computer Physics Communications} \textbf{2014}, \emph{185}, 604–613\relax
\mciteBstWouldAddEndPuncttrue
\mciteSetBstMidEndSepPunct{\mcitedefaultmidpunct}
{\mcitedefaultendpunct}{\mcitedefaultseppunct}\relax
\EndOfBibitem
\bibitem[M{\'e}ndez and Semino(2024)M{\'e}ndez, and Semino]{Mendez2024_2}
M{\'e}ndez,~E.; Semino,~R. Phase diagram of {ZIF-4} from computer simulations. \emph{J. Mater. Chem. A Mater. Energy Sustain.} \textbf{2024}, \emph{12}, 31108--31115\relax
\mciteBstWouldAddEndPuncttrue
\mciteSetBstMidEndSepPunct{\mcitedefaultmidpunct}
{\mcitedefaultendpunct}{\mcitedefaultseppunct}\relax
\EndOfBibitem
\bibitem[Perego \latin{et~al.}(2015)Perego, Salvalaglio, and Parrinello]{Perego2015}
Perego,~C.; Salvalaglio,~M.; Parrinello,~M. Molecular dynamics simulations of solutions at constant chemical potential. \emph{The Journal of Chemical Physics} \textbf{2015}, \emph{142}\relax
\mciteBstWouldAddEndPuncttrue
\mciteSetBstMidEndSepPunct{\mcitedefaultmidpunct}
{\mcitedefaultendpunct}{\mcitedefaultseppunct}\relax
\EndOfBibitem
\bibitem[Han \latin{et~al.}(2022)Han, He, Liu, Ming, Lin, Li, Zhou, and Deng]{Han2022}
Han,~J.; He,~X.; Liu,~J.; Ming,~R.; Lin,~M.; Li,~H.; Zhou,~X.; Deng,~H. Determining factors in the growth of MOF single crystals unveiled by in situ interface imaging. \emph{Chem} \textbf{2022}, \emph{8}, 1637–1657\relax
\mciteBstWouldAddEndPuncttrue
\mciteSetBstMidEndSepPunct{\mcitedefaultmidpunct}
{\mcitedefaultendpunct}{\mcitedefaultseppunct}\relax
\EndOfBibitem
\bibitem[Karmakar \latin{et~al.}(2021)Karmakar, Invernizzi, Rizzi, and Parrinello]{Karmakar2021}
Karmakar,~T.; Invernizzi,~M.; Rizzi,~V.; Parrinello,~M. Collective variables for the study of crystallisation. \emph{Molecular Physics} \textbf{2021}, \emph{119}\relax
\mciteBstWouldAddEndPuncttrue
\mciteSetBstMidEndSepPunct{\mcitedefaultmidpunct}
{\mcitedefaultendpunct}{\mcitedefaultseppunct}\relax
\EndOfBibitem
\bibitem[Piaggi and Parrinello(2019)Piaggi, and Parrinello]{Piaggi2019}
Piaggi,~P.~M.; Parrinello,~M. Calculation of phase diagrams in the multithermal-multibaric ensemble. \emph{The Journal of Chemical Physics} \textbf{2019}, \emph{150}, 244119\relax
\mciteBstWouldAddEndPuncttrue
\mciteSetBstMidEndSepPunct{\mcitedefaultmidpunct}
{\mcitedefaultendpunct}{\mcitedefaultseppunct}\relax
\EndOfBibitem
\bibitem[Behler and Parrinello(2007)Behler, and Parrinello]{Behler2007}
Behler,~J.; Parrinello,~M. Generalized Neural-Network Representation of High-Dimensional Potential-Energy Surfaces. \emph{Physical Review Letters} \textbf{2007}, \emph{98}\relax
\mciteBstWouldAddEndPuncttrue
\mciteSetBstMidEndSepPunct{\mcitedefaultmidpunct}
{\mcitedefaultendpunct}{\mcitedefaultseppunct}\relax
\EndOfBibitem
\bibitem[Méndez \latin{et~al.}(2026)Méndez, Triestram, André, Coudert, and Semino]{https://doi.org/10.48550/arxiv.2604.09084}
Méndez,~E.; Triestram,~L.; André,~D.; Coudert,~F.-X.; Semino,~R. Force Field-Agnostic Phase Classification of Zeolitic Imidazolate Framework Polymorphs. 2026; \url{https://arxiv.org/abs/2604.09084}\relax
\mciteBstWouldAddEndPuncttrue
\mciteSetBstMidEndSepPunct{\mcitedefaultmidpunct}
{\mcitedefaultendpunct}{\mcitedefaultseppunct}\relax
\EndOfBibitem
\bibitem[Sayed(2022)]{Sayed2022}
Sayed,~A.~H. \emph{Inference and Learning from Data: Learning}; Cambridge University Press, 2022\relax
\mciteBstWouldAddEndPuncttrue
\mciteSetBstMidEndSepPunct{\mcitedefaultmidpunct}
{\mcitedefaultendpunct}{\mcitedefaultseppunct}\relax
\EndOfBibitem
\bibitem[Chalaris and Samios(2000)Chalaris, and Samios]{Chalaris2000}
Chalaris,~M.; Samios,~J. Systematic molecular dynamics studies of liquid N, N-dimethylformamide using optimized rigid force fields: Investigation of the thermodynamic, structural, transport and dynamic properties. \emph{The Journal of Chemical Physics} \textbf{2000}, \emph{112}, 8581–8594\relax
\mciteBstWouldAddEndPuncttrue
\mciteSetBstMidEndSepPunct{\mcitedefaultmidpunct}
{\mcitedefaultendpunct}{\mcitedefaultseppunct}\relax
\EndOfBibitem
\bibitem[Vanommeslaeghe \latin{et~al.}(2009)Vanommeslaeghe, Hatcher, Acharya, Kundu, Zhong, Shim, Darian, Guvench, Lopes, Vorobyov, and Mackerell]{Vanommeslaeghe2009}
Vanommeslaeghe,~K.; Hatcher,~E.; Acharya,~C.; Kundu,~S.; Zhong,~S.; Shim,~J.; Darian,~E.; Guvench,~O.; Lopes,~P.; Vorobyov,~I.; Mackerell,~A.~D. CHARMM general force field: A force field for drug‐like molecules compatible with the CHARMM all‐atom additive biological force fields. \emph{Journal of Computational Chemistry} \textbf{2009}, \emph{31}, 671–690\relax
\mciteBstWouldAddEndPuncttrue
\mciteSetBstMidEndSepPunct{\mcitedefaultmidpunct}
{\mcitedefaultendpunct}{\mcitedefaultseppunct}\relax
\EndOfBibitem
\bibitem[Metropolis and Ulam(1949)Metropolis, and Ulam]{Metropolis1949}
Metropolis,~N.; Ulam,~S. The Monte Carlo Method. \emph{Journal of the American Statistical Association} \textbf{1949}, \emph{44}, 335–341\relax
\mciteBstWouldAddEndPuncttrue
\mciteSetBstMidEndSepPunct{\mcitedefaultmidpunct}
{\mcitedefaultendpunct}{\mcitedefaultseppunct}\relax
\EndOfBibitem
\bibitem[Raiteri \latin{et~al.}(2005)Raiteri, Laio, Gervasio, Micheletti, and Parrinello]{Raiteri2005}
Raiteri,~P.; Laio,~A.; Gervasio,~F.~L.; Micheletti,~C.; Parrinello,~M. Efficient Reconstruction of Complex Free Energy Landscapes by Multiple Walkers Metadynamics. \emph{The Journal of Physical Chemistry B} \textbf{2005}, \emph{110}, 3533–3539\relax
\mciteBstWouldAddEndPuncttrue
\mciteSetBstMidEndSepPunct{\mcitedefaultmidpunct}
{\mcitedefaultendpunct}{\mcitedefaultseppunct}\relax
\EndOfBibitem
\bibitem[Bussi and Tribello(2019)Bussi, and Tribello]{Bussi2019}
Bussi,~G.; Tribello,~G.~A. \emph{Biomolecular Simulations}; Springer New York, 2019; p 529–578\relax
\mciteBstWouldAddEndPuncttrue
\mciteSetBstMidEndSepPunct{\mcitedefaultmidpunct}
{\mcitedefaultendpunct}{\mcitedefaultseppunct}\relax
\EndOfBibitem
\bibitem[Sch{\"a}fer and Settanni(2020)Sch{\"a}fer, and Settanni]{Schafer2020-sd}
Sch{\"a}fer,~T.~M.; Settanni,~G. Data reweighting in metadynamics simulations. \emph{J. Chem. Theory Comput.} \textbf{2020}, \emph{16}, 2042--2052\relax
\mciteBstWouldAddEndPuncttrue
\mciteSetBstMidEndSepPunct{\mcitedefaultmidpunct}
{\mcitedefaultendpunct}{\mcitedefaultseppunct}\relax
\EndOfBibitem
\bibitem[Paszke \latin{et~al.}(2019)Paszke, Gross, Massa, Lerer, Bradbury, Chanan, Killeen, Lin, Gimelshein, Antiga, Desmaison, K\"{o}pf, Yang, DeVito, Raison, Tejani, Chilamkurthy, Steiner, Fang, Bai, and Chintala]{10.5555/3454287.3455008}
Paszke,~A. \latin{et~al.}  \emph{Proceedings of the 33rd International Conference on Neural Information Processing Systems}; Curran Associates Inc.: Red Hook, NY, USA, 2019\relax
\mciteBstWouldAddEndPuncttrue
\mciteSetBstMidEndSepPunct{\mcitedefaultmidpunct}
{\mcitedefaultendpunct}{\mcitedefaultseppunct}\relax
\EndOfBibitem
\bibitem[Murphy(2022)]{pml1Book}
Murphy,~K.~P. \emph{Probabilistic Machine Learning: An introduction}; MIT Press, 2022\relax
\mciteBstWouldAddEndPuncttrue
\mciteSetBstMidEndSepPunct{\mcitedefaultmidpunct}
{\mcitedefaultendpunct}{\mcitedefaultseppunct}\relax
\EndOfBibitem
\bibitem[Kingma and Ba(2014)Kingma, and Ba]{https://doi.org/10.48550/arxiv.1412.6980}
Kingma,~D.~P.; Ba,~J. Adam: A Method for Stochastic Optimization. 2014; \url{https://arxiv.org/abs/1412.6980}\relax
\mciteBstWouldAddEndPuncttrue
\mciteSetBstMidEndSepPunct{\mcitedefaultmidpunct}
{\mcitedefaultendpunct}{\mcitedefaultseppunct}\relax
\EndOfBibitem
\end{mcitethebibliography}

\end{document}